\documentclass[12pt]{iopart}
\usepackage{iopams,amssymb}
\usepackage{latexsym}
\usepackage{graphicx}
\usepackage{dcolumn}
\usepackage{bm,color}
\usepackage{txfonts}
\usepackage{multirow}
\usepackage[thinlines]{easytable}
\newcommand{\eq}{Eq.~}
\newcommand{\eqs}{Eqs.~}
\newcommand{\fig}{Fig.~}

\newcommand{\cf} {cf.~}
\newcommand{\ug} {\!=\!}

\newcommand{\piu} {\!+\!}
\newcommand{\meno} {\!-\!}

\newcommand{\eg} {e.g.~}

\newcommand{\rref} {Ref.~}
\newcommand{\rrefs} {Refs.~}

\begin{document}

\title{Towards computability of trace distance discord}

\author{F. Ciccarello$^1$,  T. Tufarelli$^2$, and V. Giovannetti$^3$}
\address{$^1$NEST, Istituto Nanoscienze-CNR and Dipartimento di Fisica e Chimica, Universit$\grave{a}$  degli Studi di Palermo, via Archirafi 36, I-90123 Palermo, Italy\\
$^2$ QOLS, Blackett Laboratory, Imperial College London, SW7 2BW, UK\\
$^3$ NEST, Scuola Normale Superiore and Istituto Nanoscienze-CNR, Piazza dei Cavalieri 7, I-56126 Pisa, Italy}

\date{\today}
\begin{abstract}
It is known that a reliable geometric quantifier of discord-like correlations can be built by employing the so-called trace distance. This is used to measure how far the state under investigation is from the closest ``classical-quantum" one. To date, the explicit calculation of this indicator for two qubits was accomplished only for states such that the reduced density matrix of the measured party is maximally mixed, a class that includes Bell-diagonal states. Here, we first reduce the required optimization for a general two-qubit state to the minimization of an explicit two-variable function. Using this framework, we show next that the minimum can be analytically worked out in a number of relevant cases including  quantum-classical and $X$ states. This provides an explicit and compact expression for the trace distance discord of an arbitrary state belonging to either of these important classes of density matrices. 
\end{abstract}
\maketitle

\noindent
\section{Introduction}  
The issue that the quantum correlations (QCs) of a composite state are not entirely captured by entanglement (as formerly believed) has recently emerged as a topical subject calling for the introduction of new paradigms. Despite early evidence of this problem was provided over a decade ago \cite{zurek} an impressive burst of attention to this matter has developed only in the last few years \cite{review} as witnessed, in particular, by very recent experimental works (see e.g.~\rrefs \cite{zeilinger, blatt}).
In this paper, we focus on those correlations that are associated to the notion of {\it quantum discord} \cite{zurek}. Following the introduction of this concept, a variety of different measures of QCs have been put forward (see \rref\cite{review} for a comprehensive review). A major reason behind such a proliferation of QCs indicators stems from the typical difficulty in defining a reliable measure that is easily computable. No general closed formula of quantum discord, for instance, is known (with strong indications that this is an unsolvable problem \cite{progress}) even for a pair of two-dimensional systems or ``qubits" \cite{NC}, namely the simplest composite quantum system. Unfortunately, the demand for computability typically comes at the cost of ending up with quantities that fail to be {\it bona fide} measures. In this respect, the most paradigmatic instance is embodied by the so called {\it geometric discord} (GD) \cite{dakic}, which while being effortlessly computable (and in some cases able to provide useful information) {may entail unphysical predictions}.
It can indeed grow under local operations on the unmeasured party \cite{problem-geometric}, an effect which a {physically reliable (bona fide)} indicator (\eg quantum discord) is required not to exhibit. Following an approach frequently adopted for other QCs measures, the one-sided GD is defined as the distance between the state under study and the set of classical-quantum states. The latter class features zero quantum discord with respect to the measured party, say subsystem $A$, which entails the existence of at least one set of local projective measurements on $A$ leaving the state unperturbed \cite{zurek, cc}). While the above definition in terms of a distance is clear and intuitive, it requires the use of a metric in the Hilbert space. The GD employes the Hilbert-Schmidt distance, which is defined in terms of the Schatten 2-norm. Such a distance is well-known not to fulfil the property of being contractive under trace-preserving quantum channels \cite{geometry-book,noncon}, which is indeed the reason behind the aforementioned drawback of GD \cite{tufo-note}. This naturally leads to a redefinition of the GD in terms of a metric that obeys the contractivity property. One such metric is the {\it trace distance} \cite{NC,RUSKAI}, which employs the Schatten 1-norm (or trace norm for brevity). In the remainder of this paper, we refer to the QCs geometric measure resulting from this specific choice as {\it trace distance discord} (TDD).

While investigations are still in the early stages \cite{debarba,rana,nakano,sarandy1,sarandy2}, TDD appears to enjoy attractive features{, which makes it a physically meaningful measure}.  Besides the discussed contractivity property, the trace distance is invariant under unitary transformations. More importantly, it is in one-to-one correspondence with {\it one-shot state distinguishability} \cite{nielsen}, i.e., the maximum probability to distinguish between two states through a single measurement. This operational interpretation provides evidence that the trace distance works as an accurate ``meter" in the space of quantum states {which, importantly, has a clear physical meaning.}
Another appealing advantage of TDD lies in its connection with entanglement. Recently, indeed, it was suggested to define the full amount of discord-like correlations in a system $S$ as the minimum entanglement between $S$ and the measurement apparatus created in a local measurement (see \cite{piani2,piani3,streltsov} and references therein). This way, a given entanglement measure \cite{horo} identifies a corresponding QCs indicator. Remarkably, it turns out that the latter always exceeds the entanglement between the subparts of $S$ when this is quantified via the same entanglement measure. This rigorously formalizes the idea that a composite state can feature QCs that cannot be ascribed to entanglement. In this framework, it can be shown \cite{nakano} that the entanglement counterpart of TDD is {\it negativity} \cite{negativity}, the latter being a well-known -- in general easily computable -- entanglement monotone \cite{nota-noq}.

In spite of all such interesting features, the easiness of computation of TDD in actual problems is yet to be assessed. To date, the only class of states for which a closed analytical expression has been worked out are the Bell-diagonal (BD) two-qubit states, or more generally states that appear maximally mixed to the measured party \cite{nakano,sarandy1}. While the proof of this formula is non-trivial \cite{nakano}, this does not clarify whether or not, besides its reliability, TDD brings about computability advantages as well. Owing to the high symmetry and reduced number of parameters of BD states, indeed, most if not all of the bona fide QCs measures proposed so far can be analytically calculated for this specific class \cite{frozen}.

In this paper, we take a step forward and set up the problem of the actual computation of two-qubit TDD on a new basis. We first develop a theoretical framework that reduces this task to the equivalent minimization of a two-variable explicit function, which parametrically depends on the Bloch vectors of the marginals and the singular values of the correlation matrix. Next, after re-deriving the value of TDD  for a class of density matrices that includes BD states, we discuss two further relevant cases in which the minimization problem can be analytically solved. One is the case where the correlation matrix has one non-zero singular eigenvalue, a subset of which is given by the {\it quantum-classical} states (unlike classical-quantum states these feature non-classical correlations with respect to party $A$). The other case is given by the family of {\it $X$ states} \cite{xst}, which include BD states as special cases. While these are arguably among the most studied classes of!
  two-qubit density matrices \cite{review}, the calculation of their QCs through bona fide measures is in general a demanding task. To the best of our knowledge, in particular, no closed expression for an arbitrary quantum-classical state is known to date with the exception of \rref\cite{laleh} where however an {\it ad hoc} measure exclusively devised for this specific class of states was presented. In the general case, indeed, one such state depends on four independent parameters and, moreover, features quite low symmetry. In \rrefs \cite{laleh} and \cite{creation}, for instance, closed expressions for a fidelity-based measure \cite{bruss} and the quantum discord, respectively, could be worked out only for high-symmetry two-parameter subsets of this family.

Even more involved is the calculation of QCs in the case of $X$ states, a class which depends on five independent parameters. Regarding quantum discord, an algorithm has been put forward by Ali {\it et al.} \cite{ali}. Later, however, some counterexamples of $X$ states for which such algorithm fails were highlighted \cite{counter} (see also \rref\cite{review}).

The present paper is organized as follows. In Section \ref{gen-case}, we present our method for tackling and simplifying the calculation of TDD for an arbitrary two-qubit state. This is demonstrably reduced to the minimization of an explicit two-variable function. In Section~\ref{NEWSECTION}, we apply the theory to the case of Bell states and that of density matrices having correlation matrix with uniform spectrum. 
In Section \ref{one-singular}, we show that the minimum can be analytically found in a closed form whenever the correlation matrix of the composite state features only one non-zero singular value. As an application of this finding, in Subsection \ref{QCS} we compute the TDD of the most general quantum-classical state. As a further case where the minimization in Section \ref{gen-case} can be performed explicitly, in Section \ref{x-states} we tackle the important class of $X$ states and work out the TDD for an arbitrary element of this. {In Section \ref{app}, we illustrate an application of our findings to a paradigmatic physical problem (propagation of QCs across a spin chain), where the analytical calculation of quantum discord \cite{zurek}, although possible, results in uninformative formulas. We show that, while the time behavior of TDD exhibits the same qualitative features as the quantum discord, its analytical expression is quite simple.} We finally draw our conclusions in Section \ref{concl}. A few technical details are presented in the Appendix.

\section{One-sided TDD for two-qubit states: general case}\label{gen-case}

The one-sided  TDD $\mathcal{D}^{({\rightarrow})}(\rho_{AB})$ from $A$ to $B$ of a bipartite quantum
state $\rho_{AB}$  is  defined as the 
minimal (trace norm) distance between such state and the set ${\cal CQ}$ of {\it classical-quantum} density matrices which exhibit zero quantum discord with respect to local measurements on $A$, i.e. states which admit an unravelling of the form
\begin{eqnarray} \label{CQSTATES}
\rho^{({\rightarrow})}_{AB} = \sum_j |\alpha_j\rangle_A\langle \alpha_j|\otimes \varrho_B(j)\;
\end{eqnarray}  
with  $|\alpha_j\rangle_A$ being orthonormal vectors of $A$  and  $\varrho_B(j)$ being positive (non necessarily normalized) operators of $B$.
Specifically, if $\| \Theta\|_1 = \mbox{Tr}[ \sqrt{\Theta^\dag \Theta}]$ denotes the trace norm (or Schatten 1-norm) of a generic operator $\Theta$ then
\begin{eqnarray}
\mathcal{D}^{({\rightarrow})}(\rho_{AB}) =\frac{1}{2}   \; \min_{\{\rho^{({\rightarrow})}_{AB}\}} \| \rho_{AB}-\rho^{({\rightarrow})}_{AB}\|_1 \;, \label{formaldef}
\end{eqnarray} 
the 1/2 factor  ensuring that $\mathcal{D}^{({\rightarrow})}(\rho_{AB})$ takes values between 0 and 1
[analogous definition applies  for the one-sided TDD from $B$ to $A$, $\mathcal{D}^{(\leftarrow)}(\rho_{AB})$].
The quantity in \eq(\ref{formaldef}) fulfills several requirements which make
it fit for describing non-classical correlations of the discord type~\cite{nakano}. In particular, from the properties of the trace distance~\cite{NC} it follows that $\mathcal{D}^{({\rightarrow})}(\rho_{AB})$ {\cite{nota-physical}}
\begin{enumerate}
\item[{\it i)}]   is zero if and only if $\rho_{AB}$ is one of the classical-quantum density matrices~(\ref{CQSTATES});
\item[{\it ii)}]   is invariant under the action of  an arbitrary  unitary operation $U_A\otimes V_B$ that acts locally on $A$ and $B$, i.e.  
\begin{eqnarray}
\mathcal{D}^{({\rightarrow})}(\rho_{AB}) \equiv \mathcal{D}^{({\rightarrow})}(U_A \otimes V_B\;  \rho_{AB}\; U^\dag_A \otimes V^\dag_B)\;;
\end{eqnarray}
\item[{\it iii)}]   is monotonically decreasing under completely positive and trace preserving (CPT) maps on $B$;
\item[{\it iv)}]  is an entanglement monotone  when $\rho_{AB}$ is pure.
\end{enumerate}
Furthermore, in the special case in which $A$ is a qubit Eq.~(\ref{formaldef}) can be expressed  as~\cite{nakano} 
\begin{eqnarray}\label{def1}
\mathcal{D}^{({\rightarrow})}(\rho_{AB}) =\frac{1}{2}\;   \min_{\{\Pi_A\}} \| \rho_{AB} - (\Pi_A \otimes  \mathbb{I}_B)(\rho_{AB})\|_1\;,
\end{eqnarray} 
where now the minimization is performed with respect to all possible  completely depolarizing channel $\Pi_A$ on $A$
 associated with projective measurements over an orthonormal basis, i.e.  \begin{eqnarray} \label{defpi}
 \Pi_A(\cdot\cdot\cdot) = P_A \cdot\!\cdot\!\cdot P_A \piu Q_A \cdot\!\cdot\!\cdot Q_A\;,
 \end{eqnarray}
with  $P_A\equiv |\Psi\rangle_A\langle\Psi|$ and $Q_A = \mathbb{I}_A\meno P_A$ being rank-one projectors { ($|\Psi\rangle$ is a generic one-qubit {\it pure} state)}.

In what follows, we will focus on the case where {\it both} $A$ and  $B$ are qubits. Accordingly, we parametrize 
the state $\rho_{AB}$ 
in terms of the Pauli  matrices  $\vec{\sigma}_{A(B)}\!=\!\{\sigma_{A(B)1},\sigma_{A(B)2},\sigma_{A(B)3}\}\!\equiv\!\{\sigma_{A(B)x},\sigma_{A(B)y},\sigma_{A(B)z}\}$, i.e. 
\begin{equation}
	\rho_{AB}=\frac{1}{4}( \mathbb{I}_A\otimes \mathbb{I}_B+\vec x_A\cdot\vec\sigma_A\otimes \mathbb{I}_B+ \mathbb{I}_A\otimes\vec x_B\cdot \vec\sigma_B+\!\sum_{i,j=1}^3 \; \Gamma_{ij}\sigma_{Ai}\otimes\sigma_{Bj})\label{rhoAB}
\end{equation}
where   
\begin{eqnarray}\label{blochAB}
\vec x_{A(B)}=\mbox{Tr}[ \rho_{AB} \vec{\sigma}_{A(B)}] \;,
\end{eqnarray}
is the Bloch vector corresponding to the reduced density matrix $\rho_{A(B)}$ describing the state of $A(B)$, while $\Gamma$ is the 3$\times$3 real correlation matrix given by
\begin{eqnarray}\label{Gammadef}
\Gamma_{ij}  = \mbox{Tr} [\rho_{AB}( \sigma_{Ai} \otimes \sigma_{Bj}) ]\;.
\end{eqnarray} 
Similarly, without loss of generality,  we express the orthogonal projectors $P_A$ and $Q_A$ of Eq.~(\ref{defpi}) as  
\begin{equation}
	P_A=\frac{1}{2}( \mathbb{I}_A+\hat e\cdot\vec\sigma_A)\;,\qquad Q_A=\frac{1}{2}( \mathbb{I}_A-\hat e\cdot\vec\sigma_A)\;\label{projector}
\end{equation}
with 
$\hat{e}$ being the 3-dimensional (real) unit vector associated with the pure state $|\Psi\rangle_A$ in the Bloch sphere. 
Using this and observing that 
$\Pi_A( \mathbb{I}_A)\ug \mathbb{I}_A$, 
and $\Pi_A( \vec{\upsilon} \cdot \vec{\sigma}_A) \ug (\hat{e}\cdot \vec{\upsilon})\,(\hat{e} \cdot \vec{\sigma}_A)$,  Eq.~(\ref{def1}) can be arranged as 
\begin{eqnarray}\label{def12}
\mathcal{D}^{({\rightarrow})}(\rho_{AB}) = \frac{1}{8}  \min_{\hat{e}} \|   M(\hat{e}) \|_1\;,
\end{eqnarray}
where the minimization is performed over the unit vector $\hat{e}$ 
and $M(\hat{e})$ is a $4\times 4$ matrix which admits the representation
\begin{eqnarray}
M(\hat{e}) &\ug& \label{M(e)}
  \big[ (\vec{x}_A-  (\hat{e}\cdot \vec{x}_A)  \hat{e})  \cdot  \vec{\sigma}_{A}\big]\otimes   \mathbb{I}_B \nonumber \\
  &&\qquad + \sum_{ij} \Gamma_{ij}\;\;  (\hat{x}_i - e_i \hat{e}) \cdot \vec{\sigma}_{A}  \otimes \sigma_{Bj}\;.
\end{eqnarray} 
Here, $\hat{x}_i$ is the $i$th Cartesian unit vector and $e_i \ug \hat{x}_i \!\cdot\! \hat{e}$ the $i$th component of $\hat{e}$ (note that $\sigma_{Ai}\ug\hat{x}_i \!\cdot\! \vec{\sigma}_A$).
The second term in Eq.~(\ref{M(e)}) can be further simplified by transforming $\Gamma$ into a diagonal form via its singular value decomposition~\cite{HORN}. More precisely, exploiting the
fact that  $\Gamma$ is real we can express it as 
\begin{eqnarray}
\Gamma\ug O^{\top} \;\left[ \begin{array}{ccc}
\gamma_1 & 0& 0 \\
0& \gamma_2 & 0 \\
0& 0 & \gamma_3
\end{array} \right] 
\,\Omega \;, \label{gammaDI}
\end{eqnarray}
 where $O$ and $\Omega$ 
are real orthogonal matrices of $\mbox{SO}(3)$ while $\{\gamma_i\}$ are  real (not necessarily non-negative) quantities whose moduli correspond to the 
singular eigenvalues of $\Gamma$~\cite{COMMENTO}.
We can then define the two set of vectors
\begin{eqnarray}
 \hat{w}_k =  \sum_{j=1}^3  O_{kj} \hat{x}_j\;, \qquad \hat{\upsilon}_k =  \sum_{j=1}^3   \Omega_{kj}  \hat{x}_j\;\label{wk-vk}
\end{eqnarray} 
for $k\ug1,2,3$.
As $O,\Omega\!\in\!\mbox{SO}(3)$, by construction $\{\hat{w}_k\}$ 
is an orthonormal (right-hand oriented) set of real vectors and so is $\{\hat{\upsilon}_k\}$ (each is indeed a rotation of the Cartesian unit vectors $\{ \hat{x}_j\}$). 
Using the above, we can arrange \eq(\ref{M(e)}) as
\begin{eqnarray}
M(\hat{e}) \label{dddd}
\ug  ( \vec{x}_{A\perp} \! \cdot  \vec{\sigma}_A)\otimes   \mathbb{I}_B 
\piu  \sum_{k=1}^3  \gamma_k  \left(  \vec w_{k\perp}
\cdot {\vec{\sigma}}_A\right) \otimes  \left( \hat{\upsilon}_k \cdot {\vec{\sigma}}_B\right)\;,
\end{eqnarray} 
where for compactness of notation we introduced the vectors 
 \begin{eqnarray}
 \vec{x}_{A\perp}\ug \vec{x}_A\meno (\hat{e}\cdot\vec{x}_A)  \hat{e} \;,\,\,\,\,\,\,
 \vec w_{k\perp} \ug  \hat{w}_k \meno(\hat e \cdot \hat{w}_k) \hat{e}\;
  \label{DEFWK} \end{eqnarray} 
 to represent the orthogonal component of $\vec{x}_A$ and $\hat{w}_k$  with respect to $\hat{e}$.

Note that  $\{\hat \upsilon_k\cdot\vec \sigma_B\}$ describes the transformed set of Pauli matrices under a local rotation on $B$. This set clearly fulfills all the properties of Pauli matrices as well. One can therefore redefine the $B$'s Pauli matrices as  $\{\hat \upsilon_k\cdot\vec \sigma_B\}\!\rightarrow\!\sigma_{Bk}$, which amounts to applying a local unitary on $B$. 
Let then $M'(\hat e)$ be the transformed operator obtained from $M(\hat e)$ under such rotation, i.e. 
\begin{eqnarray}
 M'(\hat e)
\ug  (\vec{x}_{A\perp}\!  \cdot  \vec{\sigma}_A) \otimes  \mathbb{I}_B+  \sum_{k=1}^3  \gamma_k\,( \vec{w}_{k\perp}\cdot {\vec{\sigma}}_A )\otimes 
\,{{\sigma}}_{Bk}\;.
\end{eqnarray} 
Since the trace norm is invariant under any local unitary we have
\begin{eqnarray} ||M(\hat{e})||_1\ug||{M}'(\hat{e})||_1 \; \label{identityM} 
\end{eqnarray} 
in line with the invariance property {\it ii)} of $\mathcal{D}^{({\rightarrow})}(\rho_{AB})$ [indeed
${M}'(\hat{e})$ is the operator~(\ref{M(e)}) associated to the state $\rho'_{AB}$ obtained from $\rho_{AB}$ via a local unitary rotation associated to the transformation $\{\hat \upsilon_k\cdot\vec \sigma_B\}\!\rightarrow\!\sigma_{Bk}$]. 
The trace norm of $M'(\hat{e})$ can now computed by diagonalizing  the operator $M'(\hat e)^\dag M'(\hat e)$. For this purpose, we recall that given two arbitrary vectors \{$\vec x$, $\vec y\,$\}, the Pauli matrices fulfil the following commutation and anti-commutation relations 
\begin{eqnarray}
[ \vec{x} \cdot \vec{\sigma}_A ,  \vec{y} \cdot \vec{\sigma}_A] &\ug& 2 i\, (\vec{x}\wedge \vec{y}) \cdot \vec{\sigma}_A\;,\label{comm} \\
\{  \vec{x} \cdot \vec{\sigma}_A \;,\;  \vec{y} \cdot \vec{\sigma}_A  \} &\ug&2\,( \vec{x}\cdot \vec{y})\;\label{anti-comm}
\end{eqnarray}
as well as the identities $\sigma_{A1}\sigma_{A2}\ug i\sigma_{A3}$, $\sigma_{A2}\sigma_{A1}\ug -i\sigma_{A3}$  and the analogous identities obtained through cyclic permutations
(in the above expression ``$\wedge$" indicates the {cross} product).
Using these, we straightforwardly end up with
\begin{eqnarray}\label{MdagM}
M'(\hat e)^\dag M'(\hat e)
&=& \left( Q+  x_{A\perp}^2 \right)    \mathbb{I} _{AB}+ \Delta +  2\;   \mathbb{I}_A \otimes  \vec{\chi}\cdot \vec{\sigma}_B,\,\,\,
\end{eqnarray} 
where ${x}_{A\perp}\ug|\vec{x}_{A\perp}|$ (throughout, $x\ug |\vec x\,|$ for any vector $\vec x$), $\vec{\chi}$ is a tridimensional real vector of components 
\begin{eqnarray} 
\chi_k &\ug& \gamma_k\;  \vec{w}_{k\perp} \cdot \vec{x}_{A\perp}\label{defchi}\;,
\end{eqnarray}
while $Q$ is a positive quantity defined as 
\begin{eqnarray} 
Q&\ug& \sum_{k=1}^3 \; \gamma_k^2 \; |\vec{w}_{k\perp}|^2\,,\label{defq}
\end{eqnarray}
and finally  $\Delta$ is the operator 
\begin{eqnarray}
\Delta &\ug& \sum_{j\neq k} \gamma_j \gamma_{k} \left[ (\vec{w}_{j\perp}\cdot \vec{\sigma}_A)( \vec{w}_{k\perp}\cdot \vec{\sigma}_A) \otimes \sigma_{Bj} \sigma_{Bk}\right]\nonumber \\  
& \ug& - 2 \gamma_1 \gamma_2 \,
(\vec{w}_{1\perp} \wedge \vec{w}_{2\perp})  \cdot \vec{\sigma}_A  \otimes \sigma_{B3}  \nonumber \\ 
&&- 2 \gamma_2 \gamma_3  \,  (\vec{w}_{2\perp} \wedge \vec{w}_{3\perp})    \cdot \vec{\sigma}_A  \otimes  \sigma_{B1} \nonumber \\
&&- 2\gamma_3 \gamma_1\;  (\vec{w}_{3\perp} \wedge \vec{w}_{1\perp})  \cdot \vec{\sigma}_A  \otimes   \sigma_{B2} \label{defdelta}\;.
\end{eqnarray} 
This expression can be simplified by observing that since the $\vec{w}_{k\perp}$'s are vectors orthogonal to $\hat{e}$ [see  Eq.~(\ref{DEFWK})] 
their mutual {cross} products must be  collinear with the latter. Indeed, introducing the spherical coordinates  $\{\theta,\phi\}$ which specify $\hat e$ in the reference frame defined by $\{\hat{w}_k\}$,
we have 
\begin{eqnarray}
(\vec w_{1\perp} \!\wedge \vec w_{2\perp})&\ug& (\hat{w}_3\cdot \hat{e})  \; \hat e = \cos\theta\; \hat{e} \;,\nonumber \\
(\vec w_{2\perp}\! \wedge \vec w_{3\perp})&\ug&(\hat{w}_1\cdot \hat{e})  \; \hat e = \sin\theta \cos\phi \; \hat{e} \;,\nonumber \\
(\vec w_{3\perp}\! \wedge \vec w_{1\perp})&\ug& (\hat{w}_2\cdot \hat{e})  \; \hat e=  \sin\theta \sin\phi  \; \hat{e}\;.
\end{eqnarray} 
Substituting these  identities in \eq(\ref{defdelta}), the operator $\Delta$ can remarkably be arranged in terms of a simple tensor product as
\begin{eqnarray} \label{delta2}
\Delta\ug-2\, (\hat{e}\cdot\vec \sigma_A) \otimes (\vec g \cdot \vec\sigma_B)\;,
\end{eqnarray} 
where $\vec{g}$ is the vector
\begin{equation}\label{defg}
\vec{g}= (\gamma_2 \gamma_3  \sin\theta \cos\phi,  \gamma_3 \gamma_1 \sin \theta \sin\phi , \gamma_1 \gamma_2 \cos\theta)\;,
\end{equation}
which is orthogonal to $\vec{\chi}$~\cite{nota-ortog}. 
Next, observe that the operator $\hat{e}\cdot\vec \sigma_A$ of Eq.~(\ref{delta2}) is Hermitian with eigenvalues $1$ and $-1$. Therefore, if $\{|0\rangle_A,|1\rangle_A\}$ are its eigenvectors we can write $\hat{e}\cdot\vec \sigma_A \ug |0\rangle_A\langle 0|\meno|1\rangle_A\langle 1|$.  Plugging this and $ \mathbb{I}_A\ug |0\rangle_A\langle 0|\piu|1\rangle_A\langle 1|$ into \eq(\ref{MdagM}) this can be arranged as 
\begin{eqnarray}\label{MdagM2}
&&M'(\hat e)^\dag M'(\hat e)
\ug  (Q\piu x_{A\perp}^2 )   \mathbb{I}_{AB}  \nonumber \\
&& \qquad \quad \piu \;\;  2 \; \left[|0 \rangle_A\!\langle 0| \otimes 
( \vec{\chi} \meno \vec{g})\!\cdot\! \vec{\sigma}_B\,   \piu
|1 \rangle_A\!\langle 1| \otimes 
(\vec{\chi} \piu \vec{g})\!\cdot\! \vec{\sigma}_B \right],\nonumber
\end{eqnarray} 
which can now be put in diagonal form. Indeed, due to the aforementioned spectrum of $\vec x\cdot\vec \sigma$, it has eigenvalues $\lambda\ug Q \piu x_{A\perp}^2 \! \pm\! 2 \sqrt{ \chi^2 \piu g^2}$, each twofold degenerate \cite{nota-ortog}.
Therefore, through Eq.~(\ref{identityM})  we end up with 
\begin{eqnarray} 
\| M(\hat{e})\|_1\ug  2\left(  \sqrt{a  + \!  \sqrt{b}}  \,+  \!  \sqrt{a  -\!\sqrt{  b}}   \; \right)\,\,,\label{mm}
\end{eqnarray} 
where
\begin{eqnarray} 
a&\ug& a(\hat{e}) \ug Q \piu x_{A\perp}^2\ug Q\piu {x}_A^2 \meno (\vec {x}_A\!\cdot\! \hat{e})^2,\label{a-def} 
\\
b&\ug&  b(\hat{e}) \ug 4 \left( \chi^2 \piu g^2\right)\label{b-def}.\,\,\,\,
\end{eqnarray} 
Note that $Q$, $x_{A\perp}$, $\chi$ and $g$ are all functions of $\hat e$ [\cf \eqs (\ref{defchi}), (\ref{defq}) and (\ref{defg})].
As $\| M(\hat{e})\|_1$ is a positive-definite function, finding its minimum is  equivalent to searching for the minimum of its square $\| M(\hat{e})\|_1^2$. Thereby 
\begin{eqnarray} \label{idenew1}
\min_{\hat e} ||M(\hat e)||_1\ug \sqrt{\,\min_{\hat e} ||M(\hat e)||_1^2}\ug \ug 2 \sqrt{2 \,\left[\min_{\hat e}h(\hat e)\right]} \;,
\end{eqnarray}
where the function $h(\hat e)$ is defined as
\begin{eqnarray}\label{defhhh}
h(\hat{e}) = a(\hat{e}) + \sqrt{a^2(\hat{e}) - b(\hat{e})}\;.
\end{eqnarray}
In conclusion, in the light of \eqs(\ref{def12}), (\ref{mm}) and (\ref{idenew1})
\begin{eqnarray}\label{def133}
\mathcal{D}^{(\rightarrow)} (\rho_{AB})&\ug& \frac{1}{4} \min_{\hat e}\left[\sqrt{a \piu  \sqrt{b}}  \piu\!  \sqrt{a  \meno\!\sqrt{  b}}\right]\nonumber 
\\ 
&\ug& \frac{1}{4}\!\sqrt{\,2\left[\min_{\hat e}h \right]}\;.\label{def13}
\end{eqnarray}
We have thus expressed our trace-norm-based measure of QCs of an arbitrary state $\rho_{AB}$ as the minimum of an {\it explicit} function of the two angles $\{\theta,\phi\}$ ($0\!\le\!\theta\!\le\!\pi$, $0\!\le\!\phi\!\le\!2\pi$). Equation~(\ref{def13}) is the first main finding of this paper. For clarity, all quantities involved in the minimization problem under investigation are pictorially represented in Fig.~\ref{figu}.

\begin{figure}
\begin{center}
\includegraphics[width=0.4\linewidth]{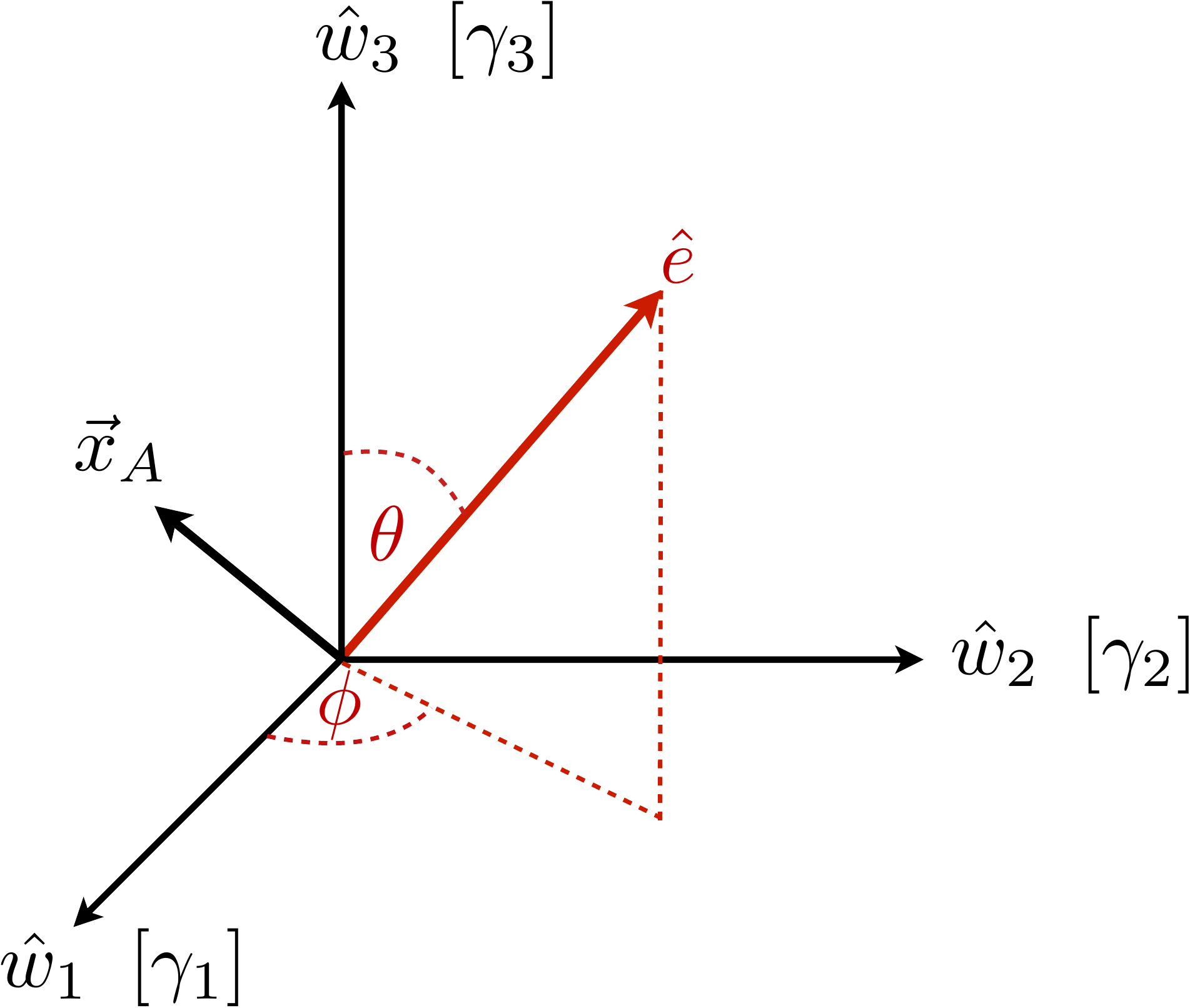}
\end{center}
\caption{(Color online) Schematics of the minimization procedure for calculating the trace distance discord $\mathcal{D}^{({\rightarrow})}(\rho_{AB})$ of a two-qubit state $\rho_{AB}$. The reference frame in which this is carried out is defined by the orthonormal set of three vectors $\{\hat w_k\}$, where each $\hat w_k$ is associated with a real singular eigenvalue $\gamma_k$ of the correlation matrix [see Eqs.~\ref{Gammadef},  \ref{gammaDI} and \ref{wk-vk}]. This frame identifies a representation for the local Bloch vector $\vec x_A$ [defined in Eq.~\ref{blochAB}]. All these quantities are drawn using solid black lines to highlight that, for a given density matrix $\rho_{AB}$, they are fixed. Instead, the unit vector $\hat{e}$ (red line) represents the direction along which a projective measurement on $A$ is performed. In the optimization procedure, $\hat e$ is varied until function $h$ in \eq(\ref{defhhh}) reaches its global minimum according to \eq(\ref{def133}).}
\end{figure}

\section{Bell diagonal states and states with homogeneous {singular values}}  \label{NEWSECTION}
The optimization in Eq.~(\ref{def13}) simplifies when the state possesses certain symmetries. 
In particular, by ordering the singular eigenvalues of $\Gamma$ as (this convention is adopted only in the present section)
\begin{eqnarray}
|\gamma_1| \geq |\gamma_2| \geq |\gamma_3|\label{ordering}\;,
\end{eqnarray}
one can show that 
\begin{eqnarray}\label{def14}
\mathcal{D}^{(\rightarrow)} (\rho_{AB}) = \frac{|\gamma_2|}{2} \;,
\end{eqnarray}
at least for two classes of states $\rho_{AB}$, which we label as `A' and `B', respectively. These are defined as
\begin{eqnarray} 
&& {\rm class\,\, A: arbitrary\,\,}  \{\gamma_k\}\,\, {\rm but}\,\, \vec{x}_A=0\nonumber\\
&&{\rm class\,\, B: arbitrary\,\,} \vec{x}_A \,\, {\rm but}\,\, \{\gamma_k\}\,\,{\rm  with\,\, the\,\, equal\,\,moduli, \, i.e.} \nonumber
\end{eqnarray}
\begin{eqnarray}
|\gamma_k|=  \gamma \; \;\; \qquad \forall k=1,2,3\;.\label{conduno}
\end{eqnarray}
We develop the proof in the following two subsections.
 
\subsection{Bell diagonal states} \label{sec:BELL}
States of class A, which include Bell diagonal states, are characterized by the property that the reduced density matrix of subsystem A is maximally mixed. For these, Eq.~(\ref{def13}) was proven in Refs.~\cite{nakano,sarandy1}
using an independent approach.  Here, we present an alternative (possibly simpler) derivation based on Eq.~(\ref{def13}). We point out that these states form a special subset of $X$ states, which we will study in full detail in Section \ref{x-states}. Here, our goal is indeed to present a straightforward application of our method for calculating the TDD developed in the previous Section.

To begin with, we observe that if $\vec{x}_A=0$ then the vector $\vec{\chi}$ in Eq.~(\ref{defchi}) vanishes, i.e. $\vec{\chi}=0$, while the function $a$ in Eq.~(\ref{a-def}) coincides with $Q$ in Eq.~(\ref{defq}).
Thereby, the function $h$ in Eq.~(\ref{defhhh}), which we have to minimize over $\hat{e}$ according to  \eq(\ref{def13}), becomes 
\begin{eqnarray}\label{defhhh11}
h = Q+ \!\sqrt{H}\; \quad{\rm { with}}\,\,\, \,  H=Q^2 - 4 g^2\;. 
\end{eqnarray}
Expressing now $Q$ in terms of $\theta$ and $\phi$ and due to the ordering in \eq(\ref{ordering}), it turns out that
\begin{eqnarray} \label{defqnew}
Q(\theta,\phi)&=& \gamma_1^2 
(1- \cos^2\phi\sin^2\theta) +  \gamma_2^2 (1- \sin^2\phi\sin^2\theta) \label{ineq1} \\ && + \; \gamma_3^2 (1-\cos^2\theta) 
\geq Q(\theta=\pi/2,\phi=0) = \gamma_2^2 + \gamma_3^2\;, \nonumber 
\end{eqnarray} 
namely $Q$ reaches its minimum value for $\theta=\pi/2$ and $\phi=0$, i.e. when $\hat{e}$ points toward $\hat{w}_1$.
 The same property holds for the function $H$. Indeed one has 
 \begin{eqnarray}
H(\theta,\phi)&=& A(\theta) \; \sin^4\phi+ B(\theta) \; \sin^2\phi +C(\theta) \nonumber \\
 &\geq& H(\theta=\pi/2,\phi=0)  \label{UNO}
=  (\gamma_2^2 - \gamma_3^2)^2\;,
\end{eqnarray} 
where we used 
\begin{eqnarray}
A(\theta)&=& \sin^4\theta \; (\gamma_1^2 - \;\gamma_2^2)^2 \geq A(0)= 0 \;,\nonumber\\
B(\theta)&=& 2 \;  \sin^2\theta\;  
(\gamma_1^2 - \;\gamma_2^2) [ (\gamma_2^2 - \;\gamma_3^2) \nonumber \\ \nonumber
&&+     \cos^2\theta\;  (\gamma_1^2 - \;\gamma_3^2)]\geq B(0)= 0\;,\nonumber \\ 
C(\theta)&=& (\gamma_2^2 - \gamma_3^2)^2+ 2 \cos^2\theta \;(\gamma_2^2 + \gamma_3^2) \;(\gamma_1^2 - \;\gamma_3^2) \nonumber\\
&&+    \cos^4\theta (\gamma_1^2 - \;\gamma_3^2)^2 \geq C(\theta=\pi/2)  \geq (\gamma_2^2 - \gamma_3^2)^2\;. \nonumber 
\end{eqnarray} 
Replacing Eq.~(\ref{ineq1}) and (\ref{UNO}) into Eq.~(\ref{defhhh11}) entails $h(\theta, \phi) \geq h(\theta=\pi/2,\phi=0) \geq  2 |\gamma_2|$, which through Eq.~(\ref{def13}) yields \eq(\ref{def14}).

\subsection{States with homogeneous $|\gamma_k|$'s} 

Class B (see definition given above) includes, for instance, mixtures of the form
$\rho_{AB} = p \varrho_A \otimes  \mathbb{I}_B/2 + (1-p) |\Psi_-\rangle_{AB} \langle \Psi_-|$ where $p\in [0,1]$, 
$\varrho_A$ is an arbitrary state of $A$, and  $|\Psi_{AB}\rangle$ is the singlet state { $|\Psi_{AB}\rangle\ug(|01\rangle_{AB}\meno|10\rangle_{AB})/\!\sqrt{2}$ [from now on, $\{|0\rangle_{A(B)},|1\rangle_{A(B)}\}$ denotes an orthonormal basis for $A$ ($B$)]}. In this case, $\gamma_k = (1-p)$ for all $k$ while $\vec{x}_A= p \vec{s}_A$ with $\vec{s}_A$ the Bloch vector of $\varrho_A$: therefore according to Eq.~(\ref{def14}) this state has a value for TDD given by $(1-p)/2$. 

To derive Eq.~(\ref{def14}), we introduce the diagonal matrix $T= \mbox{diag}(t_{11},t_{22},t_{33})$  formed by the coefficients $t_{11}, t_{22},t_{33}$ defined by the identities 
\begin{eqnarray} \label{MAPPINGT}
\gamma_j = t_{jj}\;  \gamma \;,
\end{eqnarray}
[it is clear from~(\ref{conduno}) that $t_{jj}$ can only take values $\pm 1$].
Under this condition, from Eqs.~(\ref{defchi}), (\ref{defq}) and (\ref{defg}) it then follows \begin{eqnarray}
Q &=& 2 \gamma^2 \; \nonumber \\
\vec{g} &=& \; \xi \;  \gamma^2  \; T \; \hat{e}   \Longrightarrow \;  |\vec{g}|^2 = \gamma^4 \;,
\nonumber \\ 
\vec{\chi} &=& \gamma \;  T \; \vec{x}_{A,\perp}   \Longrightarrow \;   |\vec{\chi}|^2 = \gamma^2 |\vec{x}_{A,\perp}|^2  \;, 
\end{eqnarray} 
where $\xi$ takes value either $1$ or $-1$ depending on the explicit form of the mapping~(\ref{MAPPINGT}). 
Replacing this into \eqs(\ref{a-def}), (\ref{b-def}) and (\ref{defhhh}) we end up with
\begin{eqnarray}
h = 2 \gamma^2 +2  |\vec{x}_{A,\perp}|^2  \;,
\end{eqnarray} 
which depends upon $\hat{e}$ through $|\vec{x}_{A,\perp}|^2$ only. The minimum is then achieved when $|\vec{x}_{A,\perp}|$ vanishes, which clearly occurs by taking $\hat{e}$ along the direction of $\vec{x}_A$ [recall \eq(\ref{DEFWK})]. Thus
\begin{eqnarray}
\min_{\hat e}h = 2 \gamma^2 
\end{eqnarray}
which when replaced into \eq(\ref{def13}) gives \eq(\ref{def14}), as anticipated.

\section{Correlation matrix with a single non-zero singular eigenvalue} \label{one-singular}

This class of states is important since quantum-classical states fall within it, as we show later. It is defined by [see \eq(\ref{gammaDI})]
$\gamma_2\ug\gamma_3\ug 0$
while $\gamma_1\ug\gamma$ and $\vec{x}_A$ are arbitrary (the only constraint is that the resulting $\rho_{AB}$ must be a properly defined density matrix). 
We show below that the TDD of one such state is given by
\begin{eqnarray}\label{d-1gammaDEF}
\mathcal{D}^{(\rightarrow)} (\rho_{AB}) = \frac{|\vec{\gamma}_1 \wedge \vec{x}_{A}|}{2}  \min\left\{\frac{1}{
|\vec{\gamma}_1 \pm \vec{x}_A|}\right\}\;,
\end{eqnarray}
where $\vec{\gamma}_1=|\gamma_1| \hat{w}_1$, $\hat{w}_1$ being the first element of the set $\{ \hat{w}_k\}$ defined in Eq.~(\ref{wk-vk}). \eq(\ref{d-1gammaDEF}) is another main finding of this work.

To begin with, we observe that due to $\gamma_2 \ug \gamma_3=0$ we are free to choose the direction of the Cartesian axes $\hat{w}_2$ and $\hat{w}_3$ ($\hat w_2\!\perp\!\hat w_3$) on the plane orthogonal to $\hat{w}_1$. We thus take $\hat{w}_2$ as lying on the plane formed by $\hat{w}_1$ and $\vec{x}_A$.  Hence
we can write $\vec{x}_A = \tilde{x}_{A1} \hat{w}_1 + \tilde{x}_{A2} \hat{w}_2$, where $\tilde{x}_{A1}$ and $\tilde{x}_{A2}$ are the components of 
$\vec{x}_A$  in reference frame defined by $\{ \hat{w}_k\}$. Accordingly,
 \begin{eqnarray}
 \tilde x_{A1}\ug   \hat{x}_A \!\cdot\! \hat{w}_1 = x_A \cos \alpha\;, \,\,\, \tilde{x}_{A2}\ug  \hat{x}_A \!\cdot\! \hat{w}_2 =  x_A \sin \alpha\; \label{COORD}
 \end{eqnarray}
 with $\alpha$ being the angle between $\vec x_A$ and $\hat w_1$ while $x_A\ug\! \sqrt{\tilde x_{A1}^2\piu \tilde  x_{A2}^2}$.
With the help of \eqs (\ref{defchi}), (\ref{defq}) and (\ref{defg}), in the present case $a$ and $b$ [\cf\eqs(\ref{a-def}) and (\ref{b-def})] read
\begin{eqnarray} 
a \ug\gamma^2 \piu x_A^2 \meno [\gamma^2 \tilde e_1^2 \piu (\hat{e}\cdot \vec{x}_A)^2 ]\,,\,\,\,\,
b \ug4 \gamma^2 [ \tilde x_{A1} \meno \tilde e_1 (\hat{e}\cdot \vec{x}_A)]^2, \label{ab2}\,\,\,
\end{eqnarray} 
where $ \tilde e_1 = \hat{e} \cdot \hat{w}_1$. 
Observe then that we can write
\begin{eqnarray}
a \pm \! \sqrt{b} = (\gamma \pm \tilde x_{A1})^2 + \tilde x_{A2}^2 - \left[ \gamma \tilde e_1 \pm  (\hat{e}\cdot \vec{x}_A)\right]^2\; \label{ffd111}.
\end{eqnarray} 
 It turns out that {\it both} $a \piu  \sqrt{b}$ {\it and} $a\meno \sqrt{b}$ decrease when the component of $\hat e$ on the plane formed by $\hat{w}_1$ and $\vec{x}_A$, i.e., the $\hat w_1\!-\!\hat w_2$ plane, grows. To see this, we decompose $\hat e$ as $\hat{e} \ug \vec{\varepsilon} + \vec{\varepsilon}_\perp$, where $\vec{\varepsilon} \ug \tilde e_1 \hat{w}_1 + \tilde e_2
 \hat{w}_2$ is the component of $\hat e$ on the $\hat w_1\!-\!\hat w_2$ plane, while $\vec{\varepsilon}_{\perp} \ug \tilde e_3 \hat{w}_3$ the one orthogonal to it. With these definitions, in \eq(\ref{ffd111}) we can evidently replace $\hat{e}$ with $\vec \varepsilon$ (we remind that $\tilde x_{A3}\ug0$).  Now, it should be evident that the last term of \eq(\ref{ffd111}) can be written as $- \left[ \gamma \tilde e_1 \pm  (\hat{e}\cdot \vec{x}_A)\right]^2\ug -|\varepsilon|^2[f_{\pm}(\phi,\alpha)]^2$, where $f_\pm(\phi,\alpha)\ug \gamma \cos \phi   \!\pm\!  x_A \cos (\!
 \phi \meno\alpha)$ is a function of $\phi$ (i.e., the azimuthal angle of $\hat e$) and the aforementioned $\alpha$.
Clearly, for given $\phi$ the minimum of $a\pm \sqrt{b}$ is achieved when $|\vec{\varepsilon}|$ is maximum, i.e., for $\vec{\varepsilon} \!\equiv\! \hat{e}$ or equivalently $\theta\ug\pi/2$. Thus, due to \eq(\ref{mm}), in \eq(\ref{def13}) we can safely restrict the minimization over $\hat{e}\ug(\theta,\phi)$ to the set $\hat{e}\ug(\pi/2,\phi)$.
To summarize, we need to calculate
 \begin{eqnarray} 
 \min_\phi\left[\| M(\hat{e})\|_1\Big|_{\theta=\frac{\pi}{2}}\right]\! =\!  2\!\sum_{\eta =\pm }
 \sqrt{ (\gamma +\eta  \tilde x_{A1})^2 \piu \tilde x_{A2}^2 \meno f_\eta(\phi,\alpha)^2}. \,\,\,
\label{mm221a}
\end{eqnarray} 
Through few straightforward steps (see Appendix A), $\| M(\hat{e})\|_1$ can be arranged as (we henceforth omit to specificy $\theta\ug\pi/2$)
 \begin{eqnarray} 
\| M(\hat{e})\|_1\!\ug  2 \sum_{\eta=\pm}\left| x_A \sin(\phi-\alpha) 
 \piu \eta \gamma\sin\phi\right|. \label{mm221}
\end{eqnarray} 
Exploiting the positiveness of $\| M(\hat{e})\|_1$ and the identity $(|y\piu z|\piu|y\meno z|)^2\ug4\max\{y^2,z^2\}$, where $y$ and $z$ are any two real numbers, \eq(\ref{mm221}) can be converted into
\begin{eqnarray} 
&&\| M(\hat{e})\|_1\ug 4\max \left\{\left| x_A \sin(\phi-\alpha) \right|,
 |\gamma \sin\phi|\right\} \label{mm221max} \\
&&=   4 \sqrt{x_A^2 + \gamma^2} \; \max  \left\{ |  \sin\beta \sin(\phi-\alpha) | ,  |\cos\beta \sin\phi |\right\}\;,  \nonumber 
\end{eqnarray}
where the angle $\beta$ is defined through the identity  
\begin{eqnarray}
\sin\beta = |x_A|/\sqrt{x_A^2 + \gamma^2}\;. \label{BETA}
\end{eqnarray}
Replacing $||M(\hat e)||$ so obtained into Eq.~(\ref{def12}) we can then express the one-sided TDD of our state $\rho_{AB}$ in terms of the following min-max problem,
\begin{eqnarray}
\mathcal{D}^{({\rightarrow})}(\rho_{AB}) \!\!&\ug&\!\! \frac{\sqrt{x_A^2 \piu \gamma^2}}{2} \nonumber \\ 
&& \times\!\!  \min_{\phi\in [0,2\pi]} \!\max  \left\{ |  \sin\beta \sin(\phi\meno\alpha) | ,  |\cos\beta \sin(\phi) |\right\}\!.\,\,\,\,\,\,\,\,\,\,\,\,
\end{eqnarray} 
An analytic solution is obtained by observing that the $\phi$-dependent functions $f_1(\phi)= |  \sin\beta \sin(\phi-\alpha) |$ and $f_2(\phi) =|\cos\beta \sin(\phi) |$ have the same period $\pi$ and that in the domain $\phi\!\in\![0,\pi]$ exhibit the two crossing points $\phi_{\rm c+}$ and $\phi_{\rm c-}$ given by
  \begin{eqnarray} 
  \cot(\phi_{c \pm}) = \cot \alpha \pm \left|\frac{\cot \beta}{\sin\alpha}\right|\;.
  \end{eqnarray}
By construction, the function~\eq(\ref{mm221max}) reaches is minimum either in $\phi_{c+}$ or in $\phi_{c-}$.
Therefore,
\begin{eqnarray}
\mathcal{D}^{({\rightarrow})}(\rho_{AB}) &\ug& \frac{\sqrt{x_A^2 + \gamma^2}}{2}   \min \{ |\cos\beta \sin(\phi_{\rm c+})| ,|\cos\beta \sin(\phi_{\rm c-})|\} \nonumber\\ 
&=& \frac{|\gamma \tilde x_{A2}|}{2}  \min\left\{\frac{1}{\sqrt{(\gamma\pm \tilde x_{A1})^2\!+\! \tilde x_{A2}^2}}\right\}\;, 
\end{eqnarray}
where the latter identity have been obtained through simple algebraic manipulations. To arrange this formula in a form independent of the reference frame, we make use of \eqs(\ref{COORD}) and (\ref{BETA}). This finally yields \eq(\ref{d-1gammaDEF}).

\subsection {Quantum-classical states}\label{QCS}

The result of the previous section can be exploited to provide an analytical closed formula of $\mathcal D^{(\rightarrow)}(\rho_{AB})$ for the well-known class of quantum-classical states. One such state reads
\begin{equation}
\rho_{AB}=p\, \rho_{0A}\otimes |0\rangle_B\langle 0|+(1-p)\rho_{1A}\otimes |1\rangle_B\langle 1|,\label{QC-state}
\end{equation}
where $\rho_{0(1)}$ is an arbitrary single-qubit state with associated Bloch vector $\vec s_{0(1)}$, i.e, $\rho_{0(1)}\!=\!\left( \mathbb{I}\!+\!\vec s_{0(1)}\!\cdot\!{\vec \sigma}\right)\!/2$. The state in \eq(\ref{QC-state}) represents a paradigmatic example of a separable state which is still able to feature $A\!\rightarrow\!B$ quantum correlations. On the other hand, note that the quantum discord in the opposite direction, $B\!\rightarrow\!A$, is zero by construction.

One can assume without loss of generality that $\vec s_{0}\!=\!(0,0,s_0)$ and $\vec s_{1}\!=\!(s_1 \sin \varphi,0,s_1\cos \varphi)$ with $0\!\le\!\varphi\!\le\!\pi$, i.e., the $Z$-axis of the Bloch sphere is taken along the direction of $\vec s_{0}$ while the $Y$-axis lies orthogonal to the plane containing both $\vec s_{0}$ and $\vec s_{1}$.
Vector $\vec{x}_A$ and matrix $\Gamma$ are calculated as
\begin{eqnarray} 
\vec x_A&=&\left(\,(1\meno p ) s_1 \sin\varphi,0,p s_0\piu(1\meno p) s_1 \cos\varphi\,\right),\label{xa-vec}\\
\nonumber\\
\Gamma&=&\left(\begin{array}{ccc}
0&0&(1\meno p)s_1\!\sin\varphi\\
0&0&0\\
0&0&p s_0\piu(1\meno p)s_1\!\cos\varphi
\end{array}\right).\label{Gamma-qc}
\end{eqnarray} 
$\Gamma$ has only one singular eigenvalue since its singular value decomposition yields $\gamma_2\!=\!\gamma_3\!=\!0$ and
\begin{eqnarray} 
|\gamma_1|=\gamma=\sqrt{p^2s_0^2\piu(p\meno1)s_1\left[(p\meno1)s_1\piu2p s_0\cos \varphi\right]}.\label{g1-qc}
\end{eqnarray} 
Such states therefore fall exactly in the case studied in the previous section. To apply \eq(\ref{d-1gammaDEF}), though, we  need to calculate the 
unit vectors ${\hat{w}_k}$. From the matrix \eq(\ref{Gamma-qc}), they are calculated as
\begin{eqnarray}
\hat w_1&\ug&\frac{1}{\Delta_1}\Big(\,(p\meno1) s_1 \sin \varphi,0, p s_0 \piu(p\meno1)s_1 \cos\varphi \Big),\label{w1}\\
\hat w_2&\ug&\frac{1}{\Delta_2} \left(\frac{(1\meno p) s_1 \cot\varphi\meno p s_0 \csc\varphi}{(p\meno1)s_1},0,1\right),\label{w2}\\
\hat w_3&\ug&(0,1,0),\label{w3}
\end{eqnarray}
where $\Delta_{1,2}$ are normalization coefficients. In particular, it turns out that $\Delta_1$ coincides with $\gamma$ in Eq.~(\ref{g1-qc}), i.e. $\Delta_1= \gamma$.
Hence, the vector $\vec{\gamma}_1\ug\gamma\hat w_1$ in Eq.~(\ref{d-1gammaDEF}) is given by
\begin{eqnarray}
\vec{\gamma}_1 = \Big(\,(p\meno1) s_1 \sin \varphi,0, p s_0 \piu(p\meno1)s_1 \cos\varphi \Big)\;.
\end{eqnarray} 
This, together with \eq(\ref{xa-vec}), yields the identities 
\begin{eqnarray}
|\vec{x}_A \wedge \vec{\gamma}_1| &=& 2 p (1-p) \; s_0 s_1 \sin\varphi\;, \nonumber \\
| \vec{\gamma}_1+\vec{x}_A| &=& 2 p \; s_0\;, \nonumber\\
| \vec{\gamma}_1-\vec{x}_A| &=& 2(1- p)\; s_1\;.\nonumber 
\end{eqnarray} 
Replacing these into Eq.~(\ref{d-1gammaDEF}), we end up with 
\begin{equation}
{\mathcal D}^{(\rightarrow)} (\rho_{AB})= \frac{\sin\varphi\,}{2}\,\min\{p s_{0},(1-p) s_1\},
\end{equation}
which represents the TDD of the most general quantum-classical state [\eq(\ref{QC-state})].
{This formula has a very clear interpretation in terms of the lengths of the local Bloch vectors on A, $s_0,s_1$, the angle between them $\varphi$ and the statistical weights $p,1-p$.
One can see that the maximum value of $\mathcal D^{(\rightarrow)}$ is 1/4 and is obtained for $s_0\ug s_1\ug1$, $p\ug1/2$ and $\varphi\ug\pi/2$: this corresponds to picking on system A two pure states with orthogonal Bloch vectors, that is, two vectors belonging to mutually unbiased bases. Indeed, for} these parameters, \eq(\ref{QC-state}) reduces to $\rho_{AB}\ug1/2\left(|0\rangle_A\!\langle0|\otimes |0\rangle_B\!\langle0|\piu|+\rangle_A\!\langle+|\otimes |1\rangle_B\!\langle1|\right)$ (where $|\pm\rangle\ug(|0\rangle\!\pm\!|1\rangle)/\!\sqrt{2}$), which is a paradigmatic example of separable but  quantum-correlated state. The qualitative behavior of $\mathcal D$ for $s_0\ug s_1$ and $p\ug1/2$ is fully in line with that of the quantum discord \cite{creation} and for $s_0\ug s_1\ug 1$ with that of the fidelity-based measure analyzed in \rref\cite{laleh}.

\section{$X$ states}\label{x-states}

A two-qubit $X$ state has the $X$-shaped matrix form 
\begin{equation}
\rho_{AB}=\left( \begin{array}{cccc}
\rho_{11}&0&0&\rho_{41}^*\\
0&\rho_{22}&\rho_{32}^*&0\\
0&\rho_{32}&\rho_{33}&0\\
\rho_{41}&0&0&\rho_{44}
\end{array}
\right)\label{x-state}
\end{equation}
subject to the constraints $\sum_{i=1}^4\rho_{ii}\ug1$, $\rho_{11}\rho_{44}\!\ge\! |\rho_{14}|^2$ and $\rho_{22}\rho_{33}\!\ge\! |\rho_{23}|^2$. Here, we have referred to the computational basis $\{|00\rangle_{AB},|01\rangle_{AB},|10\rangle_{AB},|11\rangle_{AB}\}$.
Without loss of generality, off-diagonal entries $\rho_{32}$ and $\rho_{41}$ can be taken as positive, i.e., $\rho_{32}\!\ge\!0$ and $\rho_{41}\!\ge\!0$ \cite{nota-offdiag}. It is straightforward to check that for such states ${x}_{A,1}\ug{x}_{A,2}\ug0$, $\,x_{A3}\ug2(\rho_{11}\piu\rho_{22})\meno1$ (that is, $\vec x_A$ lies along the $\hat{x}_3$-axis) while the correlation matrix has already the diagonal form since $\Gamma\ug{\rm diag}\left\{\gamma_1,\gamma_2,\gamma_3\right\}$ with
\begin{eqnarray}
\gamma_1\ug2 (\rho_{32}\piu\rho_{41}),\,\,\,\gamma_2\ug2 (\rho_{32}\meno\rho_{41}),\,\,\,\gamma_3\ug1\meno 2(\rho_{22}\piu\rho_{33}).\,\,\,\,\,\,\,\label{Gamma-x}
\end{eqnarray}
Hence, in the present case $\hat{w}_k\ug \hat{x}_k$ for $k\ug1,2,3$ [see \eq(\ref{wk-vk})]. We therefore have to deal with the four parameters $x_{A3}$ and $\{\gamma_k\}$. Note that the only hierarchical relation which always holds is $|\gamma_1|\!\ge\!|\gamma_2|$ [see \eq(\ref{Gamma-x})].

In what follows, we will prove that the TDD of state \eq(\ref{x-state}) is given by
\begin{eqnarray}
&& \!\!\!\!\!\!\!\!\!\!\!\!\!\!\!\!\!\!\!\!\!\!\!\!\!\!\!\!\!\!\!\!\!\!\!\!{\rm if}\, \gamma_1^2\!\meno\!\gamma_3^2\!\piu\! x_{A3}^2\!<\!0\hspace{2.2cm}\mathcal{D}^{(\rightarrow)} (\rho_{AB})\ug \frac{{|\gamma_1|}}{2}, \nonumber\\
&&\!\!\!\!\!\!\!\!\!\!\!\!\!\!\!\!\!\!\!\!\!\!\!\!\!\!\!\!\!\!\!\!\!\!\!\!{\rm if}\, \gamma_1^2\!\meno\!\gamma_3^2\!\piu\! x_{A3}^2\!\geq\!0\!\left\{\begin{array}{ll}
\!\!\!{\rm if}\, |\gamma_3|\!\geq\!|\gamma_1|,&\!\!\!\!\mathcal{D}^{(\rightarrow\!)} (\rho_{AB})\ug\frac{{|\gamma_1|}}{2}\\
\\
\!\!\!{\rm if}\, |\gamma_3|\!<\!|\gamma_1|,&\!\!\!\!\mathcal{D}^{(\rightarrow\!)} (\rho_{AB})\ug\Theta\!\left(\gamma_2^2\!\meno\!\gamma_3^2\!\piu\! x_{A3}^2\right)\!\frac{1}{2} \!
\sqrt{\frac{\gamma_1^2
\left(\gamma_2^2+x_{A3}^2\right)-\gamma_2^2 \gamma_3^2}{\gamma_1^2-\gamma_3^2+x_{A3}^2}}\! \piu \Theta\!\left[\!-\!\left(\gamma_2^2\!\meno\!\gamma_3^2\!\piu\! x_{A3}^2\right)\right]\!\frac{|\gamma_3|}{2},
\end{array}\right.\label{disc-x}\\
\nonumber
\end{eqnarray}
where we have used the Heaviside step function $\Theta(x)$ [we adopt the standard convention $\Theta(0)\ug1/2$]. 
It can be checked (see Appendix B) that for Bell-diagonal states the above expression reproduces the result of Section \ref{NEWSECTION}, i.e., the TDD is half the intermediate value among $\{|\gamma_k|\}$ [we stress that here the labelling of the $\gamma_k$'s does not imply the ordering in \eq(\ref{ordering})].

Equation~(\ref{disc-x})  can also be written in the compact form 
\begin{eqnarray}
\mathcal{D}^{(\rightarrow)} (\rho_{AB})\!=\! 
\frac{1}{2} \!\sqrt{ \frac{\gamma_1^2 \, \max\{\gamma_3^2, \gamma_2^2 + x_{A3}^2\} \!-\!\gamma_2^2 \, \min\{\gamma_3^2,\gamma_1^2\}}{
\max\{\gamma_3^2, \gamma_2^2 + x_{A3}^2\} \!-\! \min\{\gamma_3^2,\gamma_1^2\}\!+\!\gamma_1^2 \!-\!\gamma_2^2}}\;, \nonumber\\
\label{NEWEQUATION10}
\end{eqnarray}
showing that for the $X$-states the discord is only a function of the following three parameters: $|\gamma_1|$, $|\gamma_3|$ and $\sqrt{ \gamma_2^2 + x_{A3}^2}$.

To begin with, \eqs(\ref{a-def}) and (\ref{b-def}) imply that the $(\theta,\phi)$-dependent functions $a$ and $b$ entering \eq(\ref{defhhh}) [recall that $(\theta,\phi)$ specify $\hat e$] depend only on $\mu\!\equiv\!\sin^2\theta$ and $\nu\!\equiv\!\sin^2\phi$ as
\begin{eqnarray}
a\ug a_0\piu a_1 \,\mu,\label{a-powers}\,\,\,\,\,b\ug b_0\piu b_1\, \mu\piu b_2 \,\mu^2\label{ab-x},\,\,\,
\end{eqnarray}
where $\{a_i\}$ and $\{b_i\}$ are the following linear functions of $\nu$ 
\begin{eqnarray}
\!\!\!\!\!\!a_0\!\!&\ug&\gamma_1^2\piu\gamma_2^2,\,\,\,\,a_1\ug (\gamma_3^2\piu x_{A3}^2  \meno\gamma_1^2 )+ (\gamma_1^2\meno \gamma_2^2)\nu,\label{a0-a1}\\
\!\!\!\!\!\! b_0\!\!&\ug&4\gamma_1^2\gamma_2^2,\,\,\,\,\,b_2\ug4 x_{A3}^2 \left[(\gamma_3^2\meno \gamma_1^2)+(\gamma_1^2\meno \gamma_2^2)\nu
   \right],\label{b0-b2}\\
\!\!\!\!\!\! b_1\!\!&\ug&4 \left[ \gamma_2^2\gamma_3^2\piu\gamma_1^2(x_{A3}^2\meno\gamma_2^2)\piu(\gamma_1^2\meno\gamma_2^2)(\gamma_3^2\meno x_{A3}^2)\nu\right].\label{b1}
\end{eqnarray}
Clearly, $a(\mu,\nu)$ and $b(\mu,\nu)$ are defined in the square $\mathcal S$ defined by $\mathcal{S}\equiv\left\{\mu,\nu\!: 0\!\le\!\mu\!\le\!1, 0\!\le\!\nu\!\le\!1\right\}$ [and so is $h\ug a\piu\!\sqrt{a^2\meno b}$, see \eq(\ref{defhhh})].
The partial derivative of $h$ with respect to $\nu$, $\partial_\nu h$, can be arranged as $\partial_\nu{h}\ug({2 h \partial_\nu{a}\meno\partial_\nu{b}})/\left({2 \sqrt{a^2\meno b}}\right)$ (an analogous formula holds for $\partial_\mu h$). Now, due to \eqs(\ref{ab-x})-(\ref{b1}) $\partial_\nu{a}\ug(\gamma_1^2\meno\gamma_2^2)\mu$ and, notably, $\partial_\nu{b}\ug 4[\gamma_3^2\! +\! (\mu\meno 1) x_{A3}^2]\partial_\nu a$. When these are replaced in $\partial_\nu h$ we thus end up with
 \begin{eqnarray}
\partial_\nu{h}= \frac{
  h-2 \left[\gamma_3^2\! +\! x_{A3}^2(\mu-1) \right]}{ \sqrt{a^2-b}}\,\,\partial_\nu{a}.\label{derivh}
\end{eqnarray}
As witnessed by the denominator of this equation, we observe that function $h$ is in general non-differentiable at points such that $a^2\ug b$, owing to the square root $\sqrt{a^2\meno b}$ appearing in its definition, \eq(\ref{defhhh}). One then has to investigate these points carefully, as they may potentially yield extremal values of $h$ that would not be found by simply imposing $\partial_\mu h=\partial_\nu h=0$.

As a key step in our reasoning, we first demonstrate that {\it a minimum of $h$ cannot occur in the interior of $\mathcal S$}. Afterward, we minimize function $h$ on the boundary of $\mathcal S$, which will eventually lead to formula (\ref{disc-x}).

\subsection{Proof that minimum points cannot lie in the interior of $\mathcal S$}

We first address minimum points at which $h$ is differentiable, i.e., that fulfil $a^2\!\neq\! b$ entailing the existence of partial derivatives for $h$. A necessary condition for $h$ to take a minimum on these points is then $\partial_\nu h \ug0$. Based on \eq(\ref{derivh}), this can happen when either $h\ug h_0\ug 2 [\gamma_3^2\! +\! x_{A3}^2(\mu-1)]$ or $\partial_\nu a\ug 0$. 

In the latter case, as discussed above, $\partial_\nu{a}\ug(\gamma_1^2\meno\gamma_2^2)\mu$, which vanishes for $\mu\ug0$ (that is on the boundary of $\mathcal S)$ or $|\gamma_1|\ug|\gamma_2|$. Using \eq(\ref{defhhh}) and \eqs(\ref{ab-x}) through (\ref{b1}) it is easy to calculate that when $|\gamma_1|\ug|\gamma_2|$, depending on the sign of $\gamma_2^2\meno\gamma_3^2\piu x_{A3}^2$, either $h\ug2(\gamma_2^2\piu x_{A3}^2 \mu)$ or  $h\ug2[\gamma_2^2\piu (\gamma_3^2\meno \gamma_2^2)\mu]$. Thereby, in the case $|\gamma_1|\ug|\gamma_2|$ the minima of $h$ must fall on the boundary of $\mathcal S$. 

Let us now analyze the situation where $h\ug h_0\ug 2 [\gamma_3^2\! +\! x_{A3}^2(\mu-1)]$, which would also yield $\partial_\nu h\ug0$ [\cf\eq(\ref{derivh})]. As $h\ug a\piu\sqrt{a^2\meno b}$, a necessary condition for this to occur is clearly $(h_0\meno a)^2\ug a^2\meno b$. With the help of \eqs(\ref{ab-x})-(\ref{b1}), this identity can be explicitly written as $4 (1\meno\mu) \left(\gamma_1^2\meno\gamma_3^2\piu x_{A3}^2\right)\left(\gamma_2^2\meno\gamma_3^2\piu x_{A3}^2\right)\ug0$. This is fulfilled if at least one of the following identities holds: (i) $\mu\ug1$, (ii) $\gamma_1^2\meno\gamma_3^2\piu x_{A3}^2\ug0$, (iii) $\gamma_2^2\meno\gamma_3^2\piu x_{A3}^2\ug0$. Case (i) clearly corresponds to a point on the boundary of $\mathcal S$. In case (ii), using that $\gamma_2^2\meno\gamma_3^2\piu x_{A3}^2\!\le\!0$ (due to $\gamma_2^2\!\le\!\gamma_1^2\ug \gamma_3^2\meno x_{A3}^2$) we end up with $h\ug 2[(\gamma_3^2-x_{A3}^2)\piu x_{A3}^2 \mu]$. In case (iii), using that $\gamma_!
 1^2\meno\gamma_3^2\piu x_{A3}^2\!\ge\!0$
  (due to $\gamma_1^2\!\ge\!\gamma_2^2\ug \gamma_3^2\meno x_{A3}^2$) we have that $h\ug 2\{\gamma_1^2\piu[\gamma_3^2\meno\gamma_1^2\piu(\gamma_1^2\meno\gamma_3^2\piu x_{A3}^2)\nu]\mu\}$, whose minimum occurs for $\mu\ug\nu\ug0$ or $\mu\ug1$ and $\nu\ug0$ (depending on the sign of $\gamma_3^2\meno\gamma_1^2$). Hence, even in cases (ii) and (iii), the minima of $h$ fall on the boundary of $\mathcal S$. The above shows that no minima points at which $h$ is differentiable can lie in the interior $\mathcal S$.

Let us now address singular points, i.e., those at which $h$ is non-differentiable and hence minimization criteria based on partial derivatives do not apply. These points (see above discussion) are the zeros of the function $f\ug a^2\meno b$. Our aim is proving that even such points, if existing, lie on the boundary of $\mathcal S$.
Firstly, note that $f\!\ge\!0$ (we recall that $a^2\!\ge\!b$ always holds, see Section II). This means that a zero of $f$ is also a minimum point for $f$. From \eqs(\ref{ab-x})-(\ref{b1}) it is evident that $f(\mu,\nu)$ is analytic throughout the real plane. Then a necessary condition for this function to take a minimum is $\partial_\mu f\ug\partial_\nu f\ug0$. It is easy to check that $\partial_\nu f$ is a simple second-degree polynomial in $\mu$, with zeros $\mu_{{\rm s}1}\ug0$ and $\mu_{{\rm s}2}\ug(\gamma_1^2\piu\gamma_2^2\meno2\gamma_3^2\piu 2 x_{A3}^2)/[(\gamma_1^2\meno\gamma_3^2\piu x_{A3}^2)\piu(\gamma_2^2\meno\gamma_1^2)\nu]$. The former solution clearly cannot correspond to stationary points of $f$ -- in particular zeros of $f$, i.e., {\it singular points of $h$} -- that lie in the interior of $\mathcal S$ (as anticipated, a zero of $f$ is also a minimum and thus one of its stationary points).
On the other hand, by plugging $\mu_{\rm s2}$ into $\partial_\mu f$ we find  $\partial_\mu f|_{\mu=\mu_{\rm s2}}\ug 4(\gamma_1^2\meno \gamma_3^2\piu x_{A3}^2)(\gamma_2^2\meno \gamma_3^2\piu x_{A3}^2)$, which vanishes for either $\gamma_1^2\meno \gamma_3^2\piu x_{A3}^2\ug0$ or $\gamma_2^2\meno \gamma_3^2\piu x_{A3}^2\ug0$. We have already shown (see above) that in neither of these two cases $h$ can admit minima in the interior of $\mathcal{S}$.

\subsection{Minima on the boundary of $\mathcal S$}

The findings of the previous subsection show that we can restrict the search for the minimum of $h$ to the boundary of $\mathcal S$. The possible values of $h$ on the square edges corresponding to $\mu\ug0$, $\mu\ug1$, $\nu\ug0$ and $\nu\ug1$ are, respectively, given by
\begin{eqnarray}
\!\!h_{\mu=0}&\!\!\ug\!\!& \gamma_1^2\piu\gamma_2^2\piu\left|\gamma_1^2\meno\gamma_2^2\right|\ug2\gamma_1^2,\label{mu0}\\
\!\!h_{\mu=1}&\!\!\ug\!\!& \gamma_3^2 \piu x_{A3}^2\piu \gamma_2^2 \piu (\gamma_1^2\meno\gamma_2^2) \nu  \piu\left| \gamma_2^2 \meno \gamma_3^2 \piu x_{A3}^2 \piu (\gamma_1^2\meno\gamma_2^2)\nu \right|,\label{mu1}\\
\!\!h_{\nu=0}&\!\!\ug\!\!& \gamma_2^2 \piu \gamma_1^2\meno(\gamma_1^2\meno\gamma_3^2\meno x_{A3}^2) \mu \piu\!\left|\gamma_2^2\meno\gamma_1^2 (1\meno\mu) \piu(x_{A3}^2\meno\gamma_3^2)\mu\right|\!,\,\,\,\,\,\,\,\label{nu0}\\
\!\!h_{\nu=1}&\!\!\ug\!\!& \gamma_2^2 \piu \gamma_1^2\meno(\gamma_2^2\meno\gamma_3^2\meno x_{A3}^2)\mu\piu\!\left|\gamma_1^2\meno\gamma_2^2 (1\meno\mu) \piu(x_{A3}^2\meno\gamma_3^2)\mu\right|\!.\,\,\,\,\,\,\,\label{nu1}
\end{eqnarray}
From \eq(\ref{mu0}) it trivially follows that the minimum of $h$ on edge $\mu\ug0$ is given by $\min h_{\mu=0}\ug2\gamma_1^2$. In the next three dedicated paragraphs, we minimize $h$ on edges $\mu\ug1$ and $\nu\ug0,1$, respectively.

\subsubsection{Edge $\mu\ug1$}

This is the set of points $(\mu\ug1,0\!\le\!\nu\!\le\!1)$ on which function $h$ is given by \eq(\ref{mu1}). Let $h_+$ ($h_-$) be the expression taken by $h$ when the absolute value in \eq(\ref{mu1}) is positive (negative). These are easily calculated as
\begin{eqnarray}
h_+(\nu)\ug 2\left[\gamma_2^2\piu x_{A3}^2\piu(\gamma_1^2\meno\gamma_2^2)\nu\right],\,\,\,\,\,\,h_-(\nu)\ug 2\gamma_3^2.
\end{eqnarray}
Importantly, note that $h_+$ always grows with $\nu$ while $h_-$ is flat.

The argument of the absolute value [\cf\eq(\ref{mu1})] increases with $\nu$ (since $\gamma_1^2\!\ge\!\gamma_2^2$) and vanishes for $\nu\ug\nu_0\ug-(\gamma_2^2\meno\gamma_3^2\piu x_{A3}^2)/(\gamma_1^2\meno\gamma_2^2)$. Hence, it is negative (non-negative) for $\nu\!<\!\nu_0$ ($\nu\!\geq\!\nu_0$). As a consequence, $h\ug h_-$ ($h\ug h_+$) for $\nu\!<\!\nu_0$ ($\nu\!\geq\!\nu_0$). Now, the minimum of $h$ on this edge depends on the sign of $\nu_0$, which depends in turn on the sign of $\gamma_2^2\meno\gamma_3^2\piu x_{A3}^2$. Indeed, if  $\gamma_2^2\meno\gamma_3^2\piu x_{A3}^2\!<\!0$ then $\nu_0\!>\!0$ and thus $\min h_{\mu=1}\!\equiv\!\min h_-\ug2\gamma_3^2$ (recall that $h_+$ grows with $\nu$). If, instead, $\gamma_2^2\meno\gamma_3^2\piu x_{A3}^2\!\geq\!0$ then $\nu_0\!\leq\!0$ and $h\!\equiv\!h_+$ for $0\!\le\nu\!\le1$, namely throughout the edge. The minimum is thus taken at $\nu\ug0$ and reads $\min h_{\mu=1}\!\equiv\! \min h_+(\nu\ug0)\ug2(\gamma_2^2\piu x_{A3}^2)$.
To summarize,
\begin{eqnarray}
&{\rm if}&\, \gamma_2^2\meno\gamma_3^2\piu x_{A3}^2\!<\!0\,\,\,\,\,\,\min h_{\mu=1}\ug 2\gamma_3^2,\label{min-mu1-1}\\
&{\rm if}&\, \gamma_2^2\meno\gamma_3^2\piu x_{A3}^2\!\geq\!0\,\,\,\,\,\,\min h_{\mu=1}\ug 2(\gamma_2^2\piu x_{A3}^2)\label{min-mu1-2},
\end{eqnarray}

\subsubsection{Edge $\nu\ug0$}

This is the set of points ($0\!\le\!\mu\!\le\!1$, $\nu\ug0$), where $h$ is given by \eq(\ref{nu0}). Similarly to the previous paragraph, we first search for the zero of the absolute value, which is easily found as $\mu\ug\mu_0\ug(\gamma_1^2\meno\gamma_2^2)/(\gamma_1^2\meno\gamma_3^2\piu x_{A3}^2)$. Its location on the real axis fulfills
\begin{eqnarray}
&&{\rm if}\, \gamma_1^2\meno\gamma_3^2\piu x_{A3}^2\!\geq\!0\,\left\{\begin{array}{ll}
{\rm if}\, \gamma_2^2\meno\gamma_3^2\piu x_{A3}^2\!\geq\!0,\,\,\,0\!\leq\!\mu_0\!\leq\!1,\\
{\rm if}\, \gamma_2^2\meno\gamma_3^2\piu x_{A3}^2\!<\!0,\,\,\,\mu_0\!>\!1\,,\,\label{mu02}
\end{array}\right.\\
&&{\rm if}\, \gamma_1^2\meno\gamma_3^2\piu x_{A3}^2\!<\!0,\,\,\,\,\,\,\mu_0\!<\!0\,\,,\,\,\,\,\,\,\,\,\,\,\,\,\,\,\,\,\,\,\,\,\,\,\,\,\,\,\,\,\,\,\,\,\,\,\,\,\,\,\,\,\,\,\,\,\,\,\,\,\,\,\,\,\,\label{mu01}\\\nonumber
\end{eqnarray}
which we will use in our analysis.
At variance with the previous paragraph, now the absolute value in \eq(\ref{nu0}) grows (decreases) with $\mu$ for $\gamma_1^2\meno\gamma_3^2\piu x_{A3}^2\!\geq\!0$ ($\gamma_1^2\meno\gamma_3^2\piu x_{A3}^2\!<\!0$).  \eq(\ref{mu1}) straightforwardly gives
\begin{eqnarray}
h_{+}(\mu)&\!\ug\!& 2(\gamma_2^2\piu x_{A3}^2\,\mu),\,\,\,h_{-}(\mu)\ug 2\left[\gamma_1^2\piu (\gamma_{3}^2\meno \gamma_1^2)\,\mu\right],\,\,\,\,\,\,\,\,\,\,\label{hpm}
\end{eqnarray}
where  $h_{\pm}$ are defined in full analogy with the previous paragraph. Note that, while $h_+$ always grows with $\mu$, $h_-$ is an increasing (decreasing) function of $\mu$ for $|\gamma_3|\!\geq\!|\gamma_1|$ ($|\gamma_3|\!<\!|\gamma_1|$).
Let us analyze the possible situations. Based on the above, if $\gamma_1^2\meno\gamma_3^2\piu x_{A3}^2\!\geq\!0$ then $\mu_0\!\geq\!0$ and, moreover, the absolute value is negative (non-negative) for $\mu<\mu_0$ ($\mu\geq\mu_0)$. This yields $h(\mu\!<\!\mu_0)\ug h_-$ and $h(\mu\!\geq\!\mu_0)\ug h_+$. Now, two cases can occur. If $|\gamma_3|\!\geq\!|\gamma_1|$, then $h_-$ grows with $\mu$ and therefore $\min h_{\nu=0}\!\equiv\! h_{-}(\mu\ug0)\ug2\gamma_1^2$. If $|\gamma_3|\!<\!|\gamma_1|$, instead, $h_-$ decreases with $\mu$. Then the minimum of $h$ depends on whether or not $\mu_0\!\leq\!1$, which depends in turn on the sign of $\gamma_2^2\meno\gamma_3^2\piu x_{A3}^2$ according to \eq(\ref{mu02}). If $\gamma_2^2\meno\gamma_3^2\piu x_{A3}^2\!\geq\!0$ then $\mu_0\!\leq\!1$ and $h$ is minimized for $\mu\ug\mu_0$ (recall that $h_+$ always grows). This yields 
$\min h_{\nu=0}\!\equiv\! h_-(\mu_0)\ug 2\,[{\gamma_1^2
  (\gamma_2^2+x_{A3}^2)\meno\gamma_2^2 \gamma_3^2}]/({\gamma_1^2\meno\gamma_3^2\piu x_{A3}^2})$.
On the other hand, $\gamma_2^2\meno\gamma_3^2\piu x_{A3}^2\!<\!0$ implies $\mu_0\!>\!1$. Hence, $h\!\equiv\!h_-$ throughout the interval $0\!\le\!\mu\!\le\!1$ and, necessarily, $\min h_{\nu=0}\!\equiv\! h_-(\mu\ug1)\ug2\gamma_3^2$.

We are left with the case $\gamma_1^2\meno\gamma_3^2\piu x_{A3}^2\!<\!0$. In this situation, $\mu_0\!<\!0$ [\cf\eq(\ref{mu01})] and the absolute value is non-negative (negative) for $\mu\!\leq\!\mu_0$  ($\mu\!>\!\mu_0$), which gives $h\!\equiv\!h_-$ throughout this edge. Now, the analysis is simpler since, evidently, only the case $|\gamma_1|\!<\!|\gamma_3|$ is possible. Thus $h_-$ can only increase [recall \eq(\ref{hpm})] and $\min h_{\nu=0}\ug h_-(\mu\ug0)\ug2\gamma_1^2$.

To summarize, on the edge $\nu\ug0$
\begin{eqnarray}
\!\!\!\!\!\!\!\!\!\!\!\!\!\!\!\!\!\!\!\!\!\!\!\!\!\!\!\!\!\!\!\!\!\!&&{\rm if}\, \gamma_1^2\meno\gamma_3^2\piu x_{A3}^2\!<\!0\hspace{2.15cm}\min h_{\nu=0}\ug2\gamma_1^2\,\,,\,\label{min-nu0-1}\\
\!\!\!\!\!\!\!\!\!\!\!\!\!\!\!\!\!\!\!\!\!\!\!\!\!\!\!\!\!\!\!\!\!\!&&{\rm if}\, \gamma_1^2\meno\gamma_3^2\piu x_{A3}^2\!\geq\!0\,\!\!\left\{\begin{array}{ll}
\!\!\!{\rm if}\, |\gamma_3|\!\geq\!|\gamma_1|&\,\!\!\!\!\min h_{\nu=0}\ug2\gamma_1^2\,\,,\label{min-nu0-2}\\
\\
\!\!\!{\rm if}\,|\gamma_3|\!<\!|\gamma_1|&\,\!\!\!\!\min h_{\nu=0}\ug\Theta\!\left(\gamma_2^2\!\meno\!\gamma_3^2\!\piu\! x_{A3}^2\right)2\frac{\gamma_1^2
\left(\gamma_2^2\!+\!x_{A3}^2\right)\!\!-\!\!\gamma_2^2 \gamma_3^2}{\gamma_1^2-\gamma_3^2+x_{A3}^2}\piu \Theta\!\left[\!-\!\left(\gamma_2^2\!\meno\!\gamma_3^2\!\piu\! x_{A3}^2\right)\right]2\gamma_3^2
\end{array}\right.
\end{eqnarray}

\subsubsection{Edge $\nu\ug1$}

This is the set of points ($0\!\le\!\mu\!\le\!1$, $\nu\ug1$), where $h$ is given by \eq(\ref{nu1}). Similarly to the previous paragraph, we first search for the zero of the absolute value, which is easily found as $\mu\ug\mu_0\ug-(\gamma_1^2\meno\gamma_2^2)/(\gamma_2^2\meno\gamma_3^2\piu x_{A3}^2)$. Its location on the real axis fulfills
\begin{eqnarray}
&&{\rm if}\,\,\, \gamma_2^2\meno\gamma_3^2\piu x_{A3}^2\!>\!0\,\,\,\,\,\,\,\,\,\,\,\,\,\mu_0\!<\!0\,\,,\label{mu01bis}\\
&&{\rm if}\,\,\, \gamma_2^2\meno\gamma_3^2\piu x_{A3}^2\!<\!0\,\left\{\begin{array}{ll}
{\rm if}\, \gamma_1^2\meno\gamma_3^2\piu x_{A3}^2\!>\!0&\mu_0\!>\!1\\
{\rm if}\, \gamma_1^2\meno\gamma_3^2\piu x_{A3}^2\!\leq\!0&\,\,0\!\leq\!\mu_0\!\leq\!1\end{array}\right.
\end{eqnarray}
Based on \eq(\ref{nu1}), the expressions taken by $h$ on this edge when the absolute value is positive and negative are, respectively
\begin{eqnarray}
h_{+}(\mu)&\!\ug\!& 2(\gamma_1^2\piu x_{A3}^2\,\mu),\,\,\,h_{-}(\mu)\ug 2\left[\gamma_2^2\piu (\gamma_{3}^2\meno \gamma_2^2)\,\mu\right].\,\,\,\,\,\,\,\,\,\,\label{hpmbis}
\end{eqnarray}
Hence, $h_+$ always grows with $\mu$ while $h_-$ is an increasing (decreasing) function of $\mu$ for $|\gamma_3|\!\geq\!|\gamma_2|$ ($|\gamma_3|\!<\!|\gamma_2|$). We show next that, the minimum of $h$ on this edge is always given by $2\gamma_1^2$.  

Indeed, if $\gamma_2^2\meno\gamma_3^2\piu x_{A3}^2\!\geq\!0$ (implying $\gamma_1^2\meno\gamma_3^2\piu x_{A3}^2\!\geq\!0$) then $\mu_0\!\leq\!0$ and $h\!\equiv\!h_+(\mu)$ throughout the edge. The minimum is thus $\min h_{\nu=1}\ug h_+(0)\ug 2\gamma_1^2$. If, instead, $\gamma_2^2\meno\gamma_3^2\piu x_{A3}^2\!<\!0$ then $h\ug h_+$ ($h\ug h_-$) for $\mu\!\leq\!\mu_0$ ( $\mu\!>\!\mu_0$). Moreover, note that in this case one has $\gamma_3^2\!>\!\gamma_2^2$, which entails [\cf \eq (\ref{hpmbis})] that both $h_-$ and $h_+$ grow with $\mu$. Hence, the minimum is again given by $\min h_{\nu=1}\ug h_+(0)\ug 2\gamma_1^2$, which completes our proof.

\subsection{Global minimum}
To give the general expression for the minimum of $h$ it is convenient to refer to the minimization study on the edge $\nu\ug0$. Recall that the minimum of $h$ on the edges $\mu\ug0$ and $\nu\ug1$ is {\it unconditionally} given by $2\gamma_1^2$. When $\gamma_1^2\meno\gamma_3^2\piu x_{A3}^2\!<\!0$, based on \eqs(\ref{min-mu1-1}) and (\ref{min-nu0-1}) the minimum reads $\min_{\hat e}h\ug 2\gamma_1^2$ (note indeed that this case necessarily entails $\gamma_2^2\meno\gamma_3^2\piu x_{A3}^2\!<\!0$ and $\gamma_1^2\!<\!\gamma_3^2$). If instead $\gamma_1^2\meno\gamma_3^2\piu x_{A3}^2\!\geq\!0$, both signs of $\gamma_2^2\meno\gamma_3^2\piu x_{A3}^2$ as well as $|\gamma_3|\meno|\gamma_1|$ are possible. Hence, if $|\gamma_3|\!\geq\!|\gamma_1|$ upon analysis of \eqs (\ref{min-mu1-1}), (\ref{min-mu1-2}) and the first case in \eq({\ref{min-nu0-2}}) we end up with $\min_{\hat e}h\ug \min\{2\gamma_1^2,2(\gamma_2^2\piu x_{A3}^2)\}$. Let us now consider $|\gamma_3|<|\gamma_1|$. For $\gamma_2^2\meno\gamma_3^2\piu x_{A3}^2\!<\!0$, this gives $\min_{\hat e}h\ug2\gamma_3^2$. For $\gamma_2^2\meno\gamma_3^2\piu x_{A3}^2\!\geq\!0$, the global minimum is the lowest number among $2\gamma_1^2$, $2(\gamma_2^2\piu x_{A3}^2)$ and $2[{\gamma_1^2
(\gamma_2^2\piu x_{A3}^2)\meno\gamma_2^2 \gamma_3^2}]/({\gamma_1^2\meno\gamma_3^2\piu x_{A3}^2})$. Hence, to summarize,\\
\begin{eqnarray}
&&\!\!\!\!\!\!\!\!\!\!\!\!\!\!\!\!\!\!\!\!\!\!\!\!\!\!\!\!\!\!\!\!\!\!\!\!{\rm if}\, \gamma_1^2\meno\gamma_3^2\piu x_{A3}^2\!<\!0\hspace{2.15cm}\min h\ug 2\gamma_1^2, \nonumber\\
&&\!\!\!\!\!\!\!\!\!\!\!\!\!\!\!\!\!\!\!\!\!\!\!\!\!\!\!\!\!\!\!\!\!\!\!\!{\rm if}\, \gamma_1^2\meno\gamma_3^2\piu x_{A3}^2\!\geq\!0\!\left\{\begin{array}{ll}
\!\!\!{\rm if}\phantom{.}|\gamma_3|\!\geq\!|\gamma_1|&\!\!\min h\ug\Theta(\gamma_2^2\meno\gamma_3^2\piu x_{A3}^2)\,2\min\left\{\gamma_1^2,\gamma_2^2\!\piu\! x_{A3}^2\,\right\}\piu \Theta\left[\!-\!(\gamma_2^2\!\meno\!\gamma_3^2\!\piu\! x_{A3}^2)\right]2\gamma_1^2,\nonumber\\
\\
\!\!\!{\rm if}\phantom{.}|\gamma_3|\!<\!|\gamma_1|&\!\!\min h\ug\Theta\!\left(\gamma_2^2\!\meno\!\gamma_3^2\!\piu\! x_{A3}^2\right)2 \min\left\{\gamma_1^2,\gamma_2^2\piu x_{A3}^2,\!\frac{\gamma_1^2
\left(\gamma_2^2+x_{A3}^2\right)-\gamma_2^2 \gamma_3^2}{\gamma_1^2-\gamma_3^2+x_{A3}^2}\,\right\}\\
\phantom{\!\!\!{\rm if}\phantom{.}|\gamma_3|\!<\!|\gamma_1|}&\phantom{\min h\ug}\!\!\!\!\!\piu \Theta\!\left[\!-\!\left(\gamma_2^2\meno\gamma_3^2\piu x_{A3}^2\right)\right]\!2\gamma_3^2.
\end{array}\right.\\
\label{disc-x-vecchio}
\end{eqnarray}
Eq.~\ref{disc-x-vecchio} can be further simplified. Indeed, on the second line (case $\gamma_1^2\meno \gamma_3^2\piu x_{A3}^2\!\ge\!0$ and $|\gamma_3|\!\ge\!|\gamma_1|$) for $\gamma_2^2\meno \gamma_3^2\piu x_{A3}^2\!\ge\!0$ we have $\gamma_2^2\piu x_{A3}^2\!\ge\!\gamma_3^2\!\ge\!\gamma_1^2$ and therefore the minimum is $2\gamma_1^2$ regardless of $\gamma_2^2\piu x_{A3}^2\!\ge\!\gamma_3^2$. On the other hand, on the third line (case $\gamma_1^2\meno \gamma_3^2\piu x_{A3}^2\!\ge\!0$ and $|\gamma_3|\!<\!|\gamma_1|$) for $\gamma_2^2\meno \gamma_3^2\piu x_{A3}^2\!\ge\!0$ and using $\gamma_1^2\!\ge\!\gamma_2^2$ it is straightforward to prove that the rational function can never exceed both $\gamma_1^2$ and $\gamma_2^2\piu x_{A3}^2$. In light of these considerations and upon comparison of \eq(\ref{disc-x-vecchio}) with \eqs(\ref{min-nu0-1}) and (\ref{min-nu0-2}), we conclude that the global minimum of $h$ is achieved on the edge $\nu\ug0$. Therefore, using \eq(\ref{def13}) the TDD of an arbitrary two-qubit $X$ state is finally obtained as in \eq(\ref{disc-x}). Remarkably, in each case that can occur (depending on the parameters defining the state) $\mathcal{D}^{(\rightarrow)}(\rho_{AB})$ takes a relatively compact expression.

As already anticipated, for Bell-diagonal states (see Section \ref{NEWSECTION}), \eq(\ref{disc-x}) yields the result of Section \ref{sec:BELL} as shown in detail in Appendix B.

Another interesting special case occurs when in \eq(\ref{x-state}) either $\rho_{32}\ug0$ or  $\rho_{41}\ug0$. Then, due to \eq(\ref{Gamma-x}), $|\gamma_1|\!\equiv\!|\gamma_2|$ and $\sqrt{\frac{\gamma_1^2
\left(\gamma_2^2+x_{A3}^2\right)-\gamma_2^2 \gamma_3^2}{\gamma_1^2-\gamma_3^2+x_{A3}^2}}\!\rightarrow\!\frac{|\gamma_2|}{2}$. Hence, such a case always entails ${\mathcal D}^{(\rightarrow)} (\rho_{AB})\ug |\gamma_1|/2$, namely half of the absolute value of the non-zero off-diagonal entry.

\section{Application: propagation of QCs across a spin chain}\label{app}

In this Section, we present an illustrative application of our findings to a concrete problem of QCs dynamics. The problem was investigated in \rref\cite{campbell} and regards the propagation dynamics of QCs along a spin chain. Specifically, consider a chain of $N$ qubits each labeled by index $i=1,..,N$ with an associated Hamiltonian
\begin{equation}
H\ug-2 J\sum_{i=1}^{N-1}\left(\sigma_{i1}\sigma_{i+1,1}+\sigma_{i2}\sigma_{i+1,2}\right)\,.
\end{equation}
Such $XX$ model is routinely used to investigate quantum state transfer \cite{qst}.
An additional qubit, disconnected from the chain and denoted by $i\ug0$, initially shares QCs with the first qubit of the chain corresponding to $i\ug1$ (with each of the remaining qubits initially prepared in state $|0\rangle$). The problem consists in studying how the bipartite QCs between qubits 0 and $r$ with $r\ug1,...,N$ evolve in time. If $r\ug N$, in particular, one can regard this process as the end-to-end propagation of QCs across the spin chain. In \rref\cite{campbell}, the authors found a number of interesting properties, especially in comparison with the corresponding entanglement propagation. To carry out their analysis, they used the quantum discord $\mathcal D_Z^{(\rightarrow)}$\cite{zurek}. For the specific two-qubit states involved in such dynamics, $\mathcal D_Z^{(\rightarrow)}$ can be calculated analytically. Yet, this circumstance does not yield any advantage in practice since the resulting formulas are lengthy and uninformative, as pointed out by the authors themselves \cite{campbell}.
\begin{figure}
\begin{center}
\includegraphics[width=0.3\linewidth]{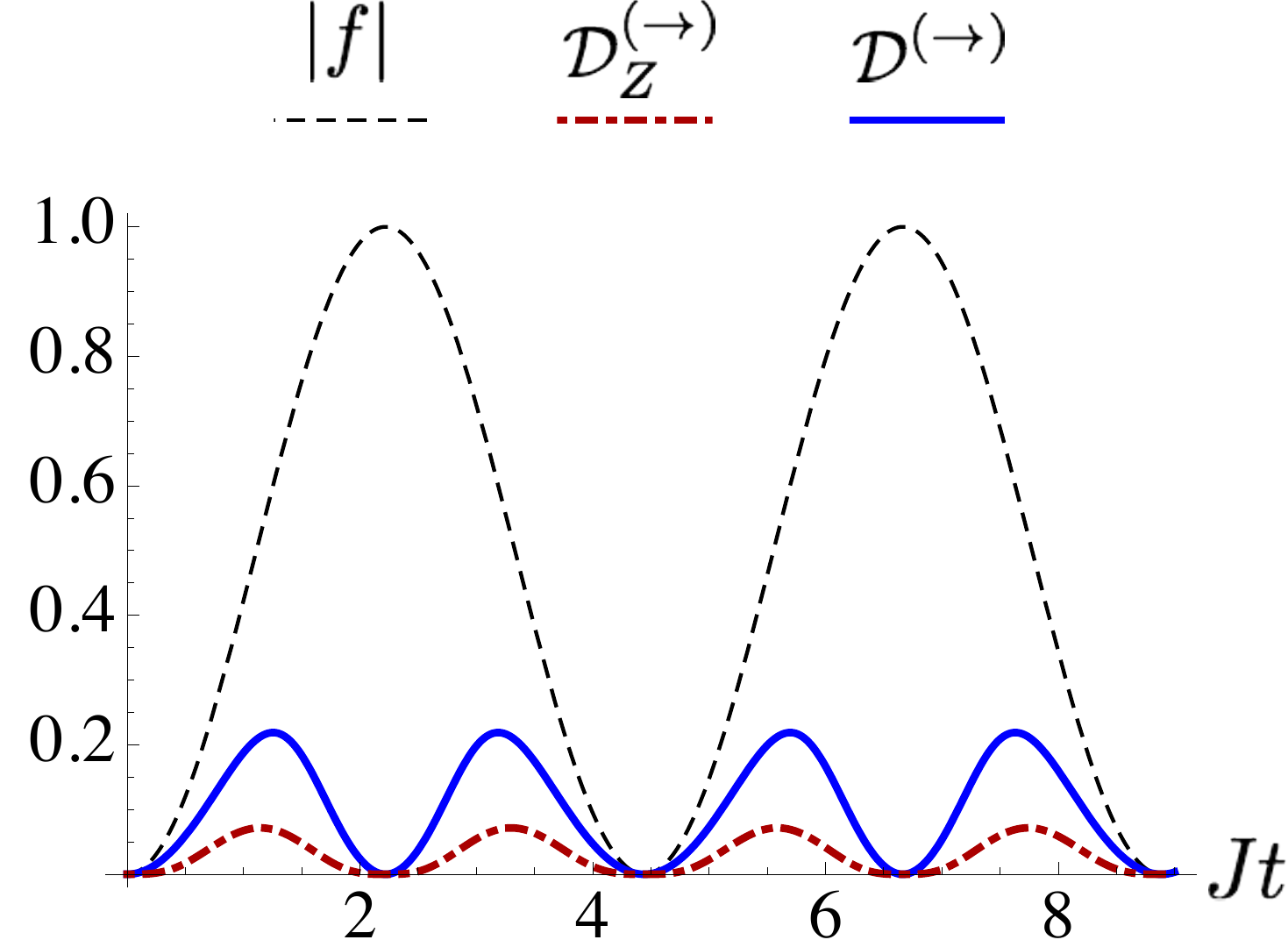},\includegraphics[width=0.3\linewidth]{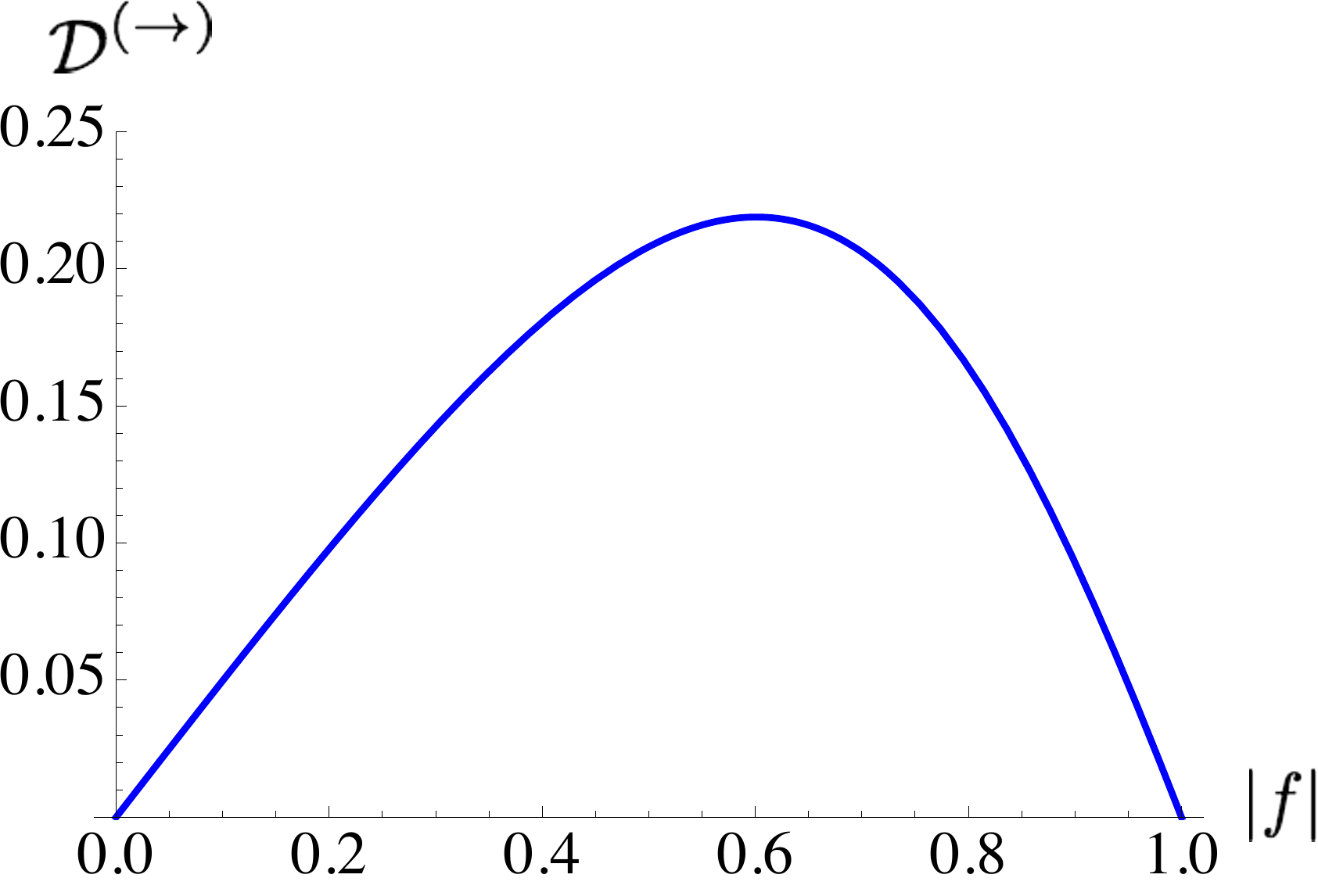}
\end{center}
\caption{(Color online) (a) Absolute value of the transfer amplitude $|f|$ (black dashed line), quantum discord $\mathcal {D}_Z^{(\rightarrow)}$ (red dot-dashed) and TDD $D^{(\rightarrow)}$ (blue solid) against time (in units of $J^{-1}$) for $N\ug3$. (b) Functional relationship between TDD and $|f|$ as given by \eq(\ref{Df}).
  \label{figu}}
\end{figure}
We next provide evidence that, if instead of $\mathcal D_Z^{(\rightarrow)}$, one uses the TDD $\mathcal D^{(\rightarrow)}$ then simple and informative formulas arise. 

It is easily demonstrated \cite{campbell} that if $\rho_{10}\ug( \mathbb{I}_{10}\piu\sigma_{11}\sigma_{01})/4$ is the initial state of qubits 1 and 0 then at time $t$ the state of $N$ and 0 reads
\begin{equation}\label{rhon0}
\rho_{N0}(t)\ug\left( \begin{array}{cccc}
\frac{2\meno|f(t)|^2}{4} & 0& 0 &\frac{f(t)}{4}\\
0&\frac{2\meno|f(t)|^2}{4} & \frac{f(t)}{4} &0\\
0& \frac{f^*(t)}{4}& \frac{|f(t)|^2}{4}&0\\
\frac{f^*(t)}{4}& 0 & 0 &\frac{|f(t)|^2}{4}\,,
\end{array} \right)
\end{equation}
where $f(t)$ is the single-excitation transition amplitude given by
\begin{equation}\label{ft}
f(t)\ug\frac{2}{N+1}\sum_{k=1}^N\sin\frac{k\pi}{N+1}\sin\frac{k\pi N}{N+1}e^{-2iJ\cos\frac{k\pi}{N+1}t}\,.
\end{equation}
Therefore, $f$ fully specifies the output state (\ref{rhon0}) and thus any corresponding QCs measure. \fig2(a) shows, in particular, the behavior of $|f(t)|$ and $\mathcal D_Z^{(\rightarrow)}[f((t)]$ for $N\ug3$, which fully reproduces the results in \rref \cite{campbell} (in absence of magnetic field). The quantum discord is evidently a non-monotonic function of $|f|$, which vanishes for $|f|\ug0,1$ exhibiting a single maximum at an intermediate value of $|f|$. There appears to be no straightforward way to prove this behavior since, as anticipated, function $\mathcal D_Z^{(\rightarrow)}(f)$ has a complicated analytical form. 

Let us now calculate $\mathcal D^{(\rightarrow)}(f)$. State \eq(\ref{rhon0}) is an $X$ state, hence our techniques of Section \ref{x-states} can be applied to calculate the corresponding TDD \cite{nota-QC}.
Using the notation of Section \ref{x-states} and observing that off-diagonal entries in \eq(\ref{rhon0}) can be replaced by their moduli (up to local unitaries that do not affect TDD) we find $\gamma_1=|f(t)|,\gamma_2=\gamma_3=0$ and $x_{A3}=1\meno |f(t)|^2$. Substituting these in Eq.\ref{NEWEQUATION10} then yields the compact expression
\begin{equation}\label{Df}
	\mathcal{D}^{(\rightarrow)} (f)\!=\! 
	\frac{1}{2}\frac{|f|\,(1-|f|^2)}{\sqrt{|f|^4-|f|^2+1}}\,,
\end{equation}
which is plotted in \fig2(b). Once $f$ is expressed as a function of time with the help of \eq(\ref{ft}) we obtain the non-monotonic time behavior of $\mathcal{D}^{(\rightarrow)}$ displayed in \fig2(a). This exhibits the same qualitative features as $\mathcal{D}_Z^{(\rightarrow)}(t)$, which shows that TDD has a predictive power analogous to the quantum discord. Unlike the latter, though, acquiring analytical insight is now straightforward. Indeed, it is immediate to see from \eq(\ref{Df}) that $\mathcal{D}^{(\rightarrow)}$ vanishes for $|f|\ug0,1$. Moreover, the equation $d \mathcal{D}^{(\rightarrow)}/d|f|\ug0$ (which is easily seen to be equivalent to an effective 3rd-degree equation) admits only one root in the range $[0,1]$ given by $|f|_{\rm M}\simeq {1}/\sqrt{3/(1\meno8/\tau\piu\tau)}\!\simeq\!0.6$ with $\tau\ug(1\piu3\sqrt{57})^{1/3}$. As $d \mathcal{D}^{(\rightarrow)}/d|f|\!>0$  for $|f|\ug0$, the TDD takes a maximum at $|f|\ug |f|_{\rm M}$ given by $\mathcal{D}^{(\rightarrow)}_{\rm M}\ug\mathcal{D}^{(\rightarrow)}(|f|_{\rm M})\!\simeq\!0.22$ [see \fig2(b)].

This paradigmatic instance illustrates the effectiveness of our findings as a tool to acquire readable and reliable informations on QCs in a concrete physical problem.

\section{Conclusions}\label{concl}

In this paper, we have addressed the issue of the computability of TDD, one of the most reliable and advantageous QCs indicators. By introducing a new method for tackling and simplifying the minimization required for its calculation in the two-qubit case, we have demonstrated that this can be reduced to the search for the minimum of an explicit two-variable function. Then, we have shown that this can be analytically found in a closed form for some relevant classes of states, which encompass arbitrary quantum-classical and $X$ states. The latter includes as a special subset the Bell diagonal states, which were the only states for which an analytical expression of TDD had been worked out prior to our work. Our results are summarized in Table~\ref{tabula}. {Finally, we have illustrated the effectiveness of our findings in a specific paradigmatic problem where, despite being achievable, the analytical calculation of quantum discord is not informative. On the contrary, TDD is readily calculated in a simple explicit form, being able at the same time to capture all the salient physical features of the QCs dynamics. Such approach could therefore prove particularly useful in order to clarify the role and physical meaning of QCs in a number of quantum coherent phenomena.}

Due to the importance of quantum-classical and $X$ states, along with the typical hindrances to the calculation of their QCs through bona fide measures, our work provides a significant contribution to the study of QCs quantifiers, by combining the desirable mathematical properties of TDD with its explicit computation for these classes of density matrices. Furthermore, we expect that the framework developed in this paper may be further exploited in future investigations to enlarge the class of quantum states that admit an analytical expression for TDD.
\begin{table}
\begin{tabular}{|m{.25\linewidth}|c|}
\hline
	\multicolumn{2}{|c|}{\bf SUMMARY OF RESULTS}\\
\hline
\hline
	\multicolumn{2}{|c|}{\bf Section~\ref{NEWSECTION}}\\
\hline
Bell diagonal states and states with uniform singular values$^*$  &$\mathcal{D}^{(\rightarrow)} (\rho_{AB}) = \frac{|\gamma_2|}{2} \;$
\\
\hline
{\textit{Example}}: \begin{footnotesize}$\rho_{AB}\!\!=\!\! p \varrho_A\boldsymbol{1}_B/2 \!+\! (1\!-\!p) |\Psi_-\rangle \langle \Psi_-|$\end{footnotesize}  & $\mathcal{D}^{(\rightarrow)} (\rho_{AB}) = \frac{1-p}{2} \;$
\\
\hline
\hline
	\multicolumn{2}{|c|}{\bf Section~\ref{one-singular}}\\
\hline
Rank-one correlation matrix [$\vec\gamma_1\equiv\gamma_1\hat w_1$] & $\mathcal{D}^{(\rightarrow)} (\rho_{AB}) = \frac{|\vec{\gamma}_1 \wedge \vec{x}_{A}|}{2}  \min\left\{\frac{1}{
|\vec{\gamma}_1 \pm \vec{x}_A|}\right\}\;,$\\
\hline
{\it Example:} QC states [Sec.~\ref{QCS}]$^\#$&${\mathcal D}^{(\rightarrow)} (\rho_{AB})= \frac{\sin\varphi\,}{2}\,\min\{p s_{0},(1-p) s_1\},$\\
\hline
\hline
	\multicolumn{2}{|c|}{\bf Section~\ref{x-states}}\\
\hline
X states$^\dagger$ & $\mathcal{D}^{(\rightarrow)} (\rho_{AB})\!=\! 
\frac{1}{2} \!\sqrt{ \frac{\gamma_1^2 \, \max\{\gamma_3^2, \gamma_2^2 + x_{A3}^2\} \!-\!\gamma_2^2 \, \min\{\gamma_3^2,\gamma_1^2\}}{
\max\{\gamma_3^2, \gamma_2^2 + x_{A3}^2\} \!-\! \min\{\gamma_3^2,\gamma_1^2\}\!+\!\gamma_1^2 \!-\!\gamma_2^2}}\;$\\
\hline
\end{tabular}
\caption{Summary of the main results of the paper. We recall that the $\gamma_j$'s indicate the (real) singular values of the correlation matrix $\Gamma$, with associated unit vectors $\hat w_j$, while $\vec x_A$ is the local Bloch vector of subsystem A, expressed in the coordinate system $\{\hat w_j\}_{j=1}^3$ --- see Section~\ref{gen-case}.\\ $^*$: In Section~\ref{NEWSECTION}, the ordering $|\gamma_1|\geq|\gamma_2|\geq|\gamma_3|$ is assumed. \\$^\#$: We recall the standard form of a Quantum-Classical state: $\rho_{AB}=p\, \rho_{0A}\otimes |0\rangle_B\langle 0|+(1-p)\rho_{1A}\otimes |1\rangle_B\langle 1|,$ where $\vec s_j$ is the Bloch vector of $\rho_{jA}$, $s_j=|\vec s_j|$ $(j=0,1)$ and $\varphi$ is the smallest angle between $\vec s_0$ and $\vec s_1$.\\ $^\dagger$: In Section~\ref{x-states}, $|\gamma_1|\geq|\gamma_2|$ is assumed, while no assumption is made on $|\gamma_3|$.\label{tabula}}
\end{table}

\section*{Acknowledgements}
We thank Adesso G for useful discussions and comments. FC and VG acknowledge support from FIRB IDEAS through project RBID08B3FM.
\bigskip
\appendix

\section{Derivation of \eq(\ref{mm221})}

We recall that $f_\pm(\phi,\alpha)\ug \gamma \cos \phi   \pm  x_A \cos (\phi \meno\alpha)$. This is a linear combination of $\cos\phi$ and $\sin \phi$, which can be arranged in terms of a single cosine as $A_{\pm}(\cos\phi\cos\delta_\pm\piu\sin\phi\sin\delta_\pm)\ug A_\pm\cos(\phi\meno\delta_{\pm})$. Using {$\tilde x_{A1}\ug x_A \cos \alpha$ [see \eq(\ref{COORD})]}, the factor is easily found as $A_{\pm}\ug\sqrt{(\gamma\!\pm\!\tilde x_{A1})^2\piu \tilde x_{A2}^2}$, while 
\begin{eqnarray}
\sin\delta_{\pm} &=& \frac{\pm x_A \sin\alpha}{\sqrt{(\gamma  \pm \tilde x_{A1})^2  +  \tilde x_{A2}^2}}=   \frac{\pm \tilde x_{A2}}{\sqrt{(\gamma  \pm \tilde x_{A1})^2  +  \tilde x_{A2}^2}}
\label{seno} \;, \\
\cos\delta_{\pm} &=& \frac{\gamma \pm x_A \cos\alpha}{\sqrt{(\gamma  \pm \tilde x_{A1})^2  +  \tilde x_{A2}^2}}= \frac{\gamma \pm \tilde x_{A1}}{\sqrt{(\gamma  \pm \tilde x_{A1})^2  +  \tilde x_{A2}^2}} \label{coseno}.
\end{eqnarray} 
Hence, $\delta_{\pm}\ug\arctan\left[\tilde x_{A2}/(\tilde x_{A1}\!\pm \!\gamma)\right]$. Therefore
\begin{eqnarray}
f_\pm(\phi,\alpha) \ug  \sqrt{(\gamma  \pm  \tilde x_{A1})^2  +  \tilde x_{A2}^2}\;  \cos(\phi -\delta_\pm).\label{f2}
 \end{eqnarray} 

Replacing \eq(\ref{f2}) into \eq(\ref{mm221a}) of the main text yields
 \begin{eqnarray} 
 \| M(\hat{e})\|_1\!\! \!&\ug& \! \!2\!\sum_{\eta =\pm} \! \sqrt{\left[ (\gamma \piu\eta  \tilde x_{A1})^2 \piu \tilde x_{A2}^2 \right]\left[1 \meno   \cos^2(\phi\meno  \delta_\eta)\right]} \, \nonumber \\  
 \! \!&\ug& \! \!2\!\sum_{\eta =\pm}  \!\sqrt{ (\gamma \piu\eta  \tilde x_{A1})^2 \piu  \tilde x_{A2}^2}  \,\,|\sin(\phi\meno  \delta_\eta)|  \nonumber \\
  \! \!&\ug& \! \!2\!\sum_{\eta =\pm} \! \sqrt{ (\gamma \piu \eta  \tilde x_{A1})^2 \piu  \tilde x_{A2}^2} \; |\sin\phi \cos \delta_\eta \meno \cos\phi \sin \delta_\eta| \nonumber.
\end{eqnarray} 
Eliminating now $\sin\delta_\pm$ and $\cos\delta_\pm$ through \eqs(\ref{seno}) and (\ref{coseno}) we end up with \eq(\ref{mm221}) of the main text.

\section{\eq(\ref{disc-x}) for Bell diagonal states}

Bell diagonal states are defined as a mixture of the four Bell states. This immediately yields that they fulfil $\vec x_A\ug\vec x_B\ug0$, that is, the reduced density matrix describing the state of either party is maximally mixed. Therefore, the corresponding density matrix can be expanded as a linear combination of $ \mathbb{I}_A\!\otimes\! \mathbb{I}_B$ and $\{\sigma_{Ak}\!\otimes\!\sigma_{Bk}\}$. As each of these four operators has an $X$-form matrix representation [\cf\eq(\ref{x-state})] Bell-diagonal states are $X$ states. Hence, in \eq(\ref{disc-x}) $\gamma_{i}^2\meno\gamma_3^2\piu x_{A3}^2\!\rightarrow\!\gamma_{i}^2\meno\gamma_3^2$ for $i\ug1,2$. In this case, the square root in \eq(\ref{disc-x}) coincides with $|\gamma_2|$ and the TDD reduces to
\begin{eqnarray}
|\gamma_3|\!\geq\!|\gamma_1|&\Rightarrow&{\mathcal D}^{(\rightarrow)} (\rho_{AB})\ug \frac{{|\gamma_1|}}{2}, \nonumber\\
 |\gamma_3|\!<\!|\gamma_1|&\Rightarrow&{\mathcal D}^{(\rightarrow)}(\rho_{AB})\ug\frac12\max\{|\gamma_2|,|\gamma_3|\}
\end{eqnarray}
It is immediate to check that the above is equivalent to state that $\mathcal{D}^{(\rightarrow)} (\rho_{AB})$ is half of the {\it intermediate} value among $\{|\gamma_1|,|\gamma_2|,|\gamma_3|\}$, which fully agrees with \rrefs\cite{nakano,sarandy1} and the findings of Sec.~\ref{sec:BELL}.


\begin {thebibliography}{99}
\bibitem{review} Modi K, Brodutch A, Cable H, Paterek T, and Vedral V 2012 {\it Rev. Mod. Phys.} {\bf 84}, 1655.
\bibitem{zurek} Ollivier H and Zurek W H 2001 {\it Phys. Rev. Lett.} {\bf 88}, 017901; Henderson L and Vedral V 2001 {\it J. Phys. A} {\bf 34}, 6899.
\bibitem{zeilinger} Dakic B, Lipp Y O, Ma X, Ringbauer M, Kropatschek S, Barz S, Paterek T, Vedral V, Zeilinger A, Brukner C 2012 {\it Nat. Phys.} {\bf 8}, 666.
\bibitem{blatt} Lanyon B P, Jurcevic P,Hempel C, Gessner M, Vedral V, Blatt R, Roos C F (2013) {\it arXiv:1304.3632}.
\bibitem{NC} Nielsen M A and Chuang I L 2000, \textit{Quantum Computation and Quantum Information} (Cambridge University Press, Cambridge, U. K.).
\bibitem{progress} Girolami D and Adesso G 2011 {\it Phys. Rev. A} {\bf 83}, 052108.
\bibitem{dakic} Dakic B, Vedral V, and Brukner C 2010 {\it Phys. Rev. Lett.} {\bf 105}, 190502.
\bibitem{problem-geometric} Piani M 2012 {\it Phys. Rev. A} {\bf 86}, 034101; Hu X, Fan H, Zhou D L, and Liu W-M 2012 {\it arXiv:1203.6149}.
\bibitem{cc}  Groisman B, Kenigsberg D, and Mor T 2007, {\it arXiv:quant-ph/0703103}.
\bibitem{geometry-book} Bengtsson I and Zyczkowski K 2006 {\it Geometry of Quantum States: An Introduction to Quantum Entanglement}, (Cambridge University Press, Cambridge, U. K.). 
{\bibitem{noncon} Ozawa M {\it Phys. Lett. A} 2000 {\bf 268}, 158;
Wang X and Schirmer S G 2009 {\it Phys. Rev. A}
{\bf 79}, 052326.}
\bibitem{tufo-note} There is an {\it ad hoc} alternative strategy to overcome some of the pathologies of GD by rescaling this in terms of the state purity, as shown in Tufarelli T, MacLean T, Girolami D, Vasile R, and Adesso G 2013 {\it J. Phys A.} {\bf 46}, 275308. Although this results in a more reliable QCs quantifier, it is still not sufficient to cure the lack of contractivity of GD.
\bibitem{RUSKAI}
Ruskai M B 1994 {\it Rev. Math. Phys.} {\bf 6}, 1147.
\bibitem{debarba} Debarba T, Maciel T O, and Vianna R O 2012 {\it Phys. Rev. A} {\bf 86},
024302, see also Erratum at arXiv:1207.1298v3 (2012).
\bibitem{rana} Rana S and Parashar P 2013 {\it Phys. Rev. A} {\bf 87}, 016301.
\bibitem{nakano} Nakano T, Piani M, and Adesso G 2013 {\it Phys. Rev. A} {\bf 88}, 012117.
\bibitem{sarandy1} Paula F M, de Oliveira T R, and Sarandy M S 2013 {\it Phys. Rev. A} {\bf 87}, 064101.
\bibitem{sarandy2} Montealegre J D, Paula F M, Saguia A, and Sarandy M S 2013{\it Phys. Rev. A} {\bf 87}, 042115.
\bibitem{nielsen} Gilchrist A, Langford N K and Nielsen M A 2005 {\it Phys. Rev. A} {\bf 71}, 062310.
\bibitem{piani2} Piani M and Adesso G 2012 {\it Phys. Rev. A} {\bf 85}, 040301(R). 
\bibitem{piani3} Piani M, Gharibian S, Adesso G, Casalmiglia J, 
Horodecki P, and Winter A 2011 {\it Phys. Rev. Lett.} {\bf106}, 220403.
\bibitem{streltsov} Streltsov A, Kampermann H, and Bru\ss D 2011 {\it Phys. Rev. Lett.} {\bf106}, 160401. 
\bibitem{horo} Horodecki R, Horodecki P, Horodecki M, and Horodecki K 2009 {\it Rev. Mod. Phys.} {\bf 81}, 865.
\bibitem{negativity} Zyczkowski K, Horodecki P, Sanpera A, and Lewenstein M 1998 {\it Phys. Rev. A} {\bf 58}, 883; Vidal G and Werner R F 2002 {\it Phys. Rev. A} {\bf 65}, 032314.
\bibitem{nota-noq} In \rref\cite{nakano}, such a counterpart is termed {\it partial negativity of quantumness}. Rigorously speaking, if the TDD is defined as the trace distance from the closest classical state then it coincides with the negativity of quantumness when the measured party is a qubit. This is sufficient for our goals since we will deal with two-qubit states throughout.
\bibitem{xst} Yu T and Eberly J H 2007 {\it Quant. Inf. Comp.} {\bf 7}, 459; Rau A R P 2009 {\it J. Phys. A: Math. Theor.} {\bf 42}, 412002; Quesada N, Qasimi A, and James D F V 2012 {\it J. Mod. Opt.} {\bf 59}, 1322.
\bibitem{frozen} Aaronson B, Lo Franco R, and Adesso G 2013 {\it arXiv:1304.1163}.
\bibitem{laleh} Abad T, Karimipour V, and Memarzadeh L 2012 {\it Phys. Rev. A} {\bf 86}, 062316.
\bibitem{creation} Ciccarello F and Giovannetti V 2012 {\it Phys. Rev. A} {\bf 85}, 010102(R).
\bibitem{bruss} Streltsov A, Kampermann H, and Bruss D 2011 {\it Phys. Rev. Lett.}
{\bf 107}, 170502.
\bibitem{ali} Ali M, Rau A R P, and Alber G 2010 {\it Phys. Rev. A} {\bf 81}, 042105.
\bibitem{counter} Lu X-M, Ma J, Xi Z, and Wang X 2011 {\it Phys. Rev. A} {\bf 83},
012327; Chen Q, Zhang C, Yu S, Yi X X, and Oh C H 2011 {\it Phys. Rev. A} {\bf 84}, 042313.
{\bibitem{nota-physical} We stress that these properties should not be regarded as mere mathematical features.  On the contrary, they embody the requirement that the used measure fulfil some fundamental physical constraints.}
\bibitem{HORN} Horn R A and Johnson C R 1990, {\it Matrix Analysis} (Cambridge University Press, Cambridge, U. K.).
\bibitem{COMMENTO} Dealing with $O$, $\Omega$ that are elements of $\mbox{SO}(3)$ --  instead of its subset $\mbox{O}(3)$ -- is fundamental to ensure that the orthonormal sets of vectors pin \eq(\ref{wk-vk}) are properly right-hand oriented.  This  possibility
comes explicitly from the fact that we allow for negative $\gamma_k$'s in Eq.~(\ref{gammaDI}). 
Indeed, the standard singular value decomposition would yield
$\Gamma\ug \tilde{O}^{\top} {\rm diag}(|\gamma_1|,|\gamma_2|,|\gamma_3|) \tilde{\Omega}$
with $\tilde{O},\tilde{\Omega}$ elements of $O(3)$~\cite{HORN}. The last identity can then put in the form \eq(\ref{gammaDI}) by
observing that there exist $T, T' $ diagonal elements of $\mbox{O}(3)$ representing spatial inversions  and $O, \Omega \in \mbox{SO}(3)$  which allow us to write  
$\tilde{O} =  T O$ and $\tilde{\Omega} = T' \Omega$. Accordingly, we obtain 
$\Gamma\ug {O}^{\top} {\rm diag} \; T (|\gamma_1|,|\gamma_2|,|\gamma_3|) T' {\Omega}$, which coincides with 
\eq(\ref{gammaDI}) once one observes that by construction the diagonal entries of $T$ and $T'$ can only be equal to either $1$ or $-1$.
\bibitem{nota-ortog}  The orthogonality between $\vec{g}$ of Eq.~(\ref{defg}) and $\vec{\chi}$ of Eq.~(\ref{defchi})  follows from the fact 
that when computing $\vec{\chi} \cdot \vec{g}$ 
the $\gamma$'s can be factored out of the sum  to give $\vec{\chi} \cdot \vec{g} \ug \gamma_1\gamma_2\gamma_3  \Big[ \vec{x}_{A} - \hat{e} (\hat{e}\cdot \vec{x}_A)\Big]  \cdot\hat{e} \ug0$.
Exploiting this identity one can then write   $|\vec{\chi} \!\pm\! \vec{g}|\!\equiv\!\sqrt{\chi^2\piu g^2}$. 
\bibitem{nota-offdiag} {One can indeed get rid of phase factors $e^{i\arg\rho_{32}}$ and $e^{i\arg\rho_{41}}$ through local unitaries, which does not affect $\mathcal{D}^{(\rightarrow)}(\rho_{AB})$}.
\bibitem{campbell} Campbell S, Apollaro T J G, Di Franco C, Banchi L, Cuccoli A, Vaia R, Plastina F, and Paternostro M 2011 {\it Phys. Rev. A} {\bf 84}, 052316.
\bibitem{qst} Bose S 2007 {\it Contemp. Phys.} {\bf 48}, 13; Apollaro T J G, Lorenzo S, and Plastina F 2013 {\it Int. J. Mod. Phys. B} {\bf 27}, 1345035.
\bibitem{nota-QC} It can be shown that in line with \rref\cite{campbell} state (\ref{rhon0}) is quantum-classical too, hence the formulas in Section \ref{QCS} can also be used. It is however more immediate to use those in Section \ref{x-states}.
\end {thebibliography}
\end{document}